\documentclass[12pt,preprint]{aastex}
\usepackage{natbib,graphics,graphicx,multirow}

\slugcomment{\today}

\shorttitle{Quasar Classification from Optical and Mid-IR Photometry}
\shortauthors{Richards et al.}

\begin{document}

\title{Bayesian High-Redshift Quasar Classification from Optical and Mid-IR Photometry}

\author{
Gordon T. Richards,\altaffilmark{1,2}
Adam D. Myers,\altaffilmark{2,3}
Christina M. Peters,\altaffilmark{1}
Coleman M. Krawczyk,\altaffilmark{1}
Greg Chase,\altaffilmark{1}
Nicholas P. Ross,\altaffilmark{1}
Xiaohui Fan,\altaffilmark{4}
Linhua Jiang,\altaffilmark{5}
Mark Lacy,\altaffilmark{6}
Ian D. McGreer,\altaffilmark{4}
Jonathan R. Trump,\altaffilmark{7}
and
Ryan N. Riegel\altaffilmark{8}
}


\altaffiltext{1}{Department of Physics, Drexel University, 3141 Chestnut Street, Philadelphia, PA 19104, USA.}
\altaffiltext{2}{Max Planck Institut f\"{u}r Astronomie, K\"{o}nigstuhl 17, Heidelberg, Germany 69117.}
\altaffiltext{3}{Department of Physics and Astronomy, University of Wyoming, Laramie, WY 82071, USA.}
\altaffiltext{4}{Steward Observatory, University of Arizona, Tucson, AZ 85721, USA.}
\altaffiltext{5}{Kavli Institute for Astronomy and Astrophysics, Peking University, Beijing 100871, P.\ R.\ China.}
\altaffiltext{6}{National Radio Astronomy Observatory, 520 Edgemont Road, Charlottesville, VA 22903, USA.}
\altaffiltext{7}{Department of Astronomy and Astrophysics, 525 Davey Lab, The Pennsylvania State University, University Park, PA 16802, USA.}
\altaffiltext{8}{Skytree, Inc., 1731 Technology Drive, Suite 700, San Jose, CA 95110, USA.}

\begin{abstract}

  We identify 885,503 type 1 quasar candidates to $i\lesssim22$ using
  the combination of optical and mid-IR photometry.  Optical
  photometry is taken from the Sloan Digital Sky Survey-III: Baryon
  Oscillation Spectroscopic Survey (SDSS-III/BOSS), while mid-IR
  photometry comes from a combination of data from the Wide-Field
  Infrared Survey Explorer ({\em WISE}) ``ALLWISE'' data release and
  several large-area {\em Spitzer Space Telescope} fields.  Selection
  is based on a Bayesian kernel density algorithm with a training
  sample of 157,701 spectroscopically-confirmed type-1 quasars with
  both optical and mid-IR data.  Of the quasar candidates, 733,713
  lack spectroscopic confirmation (and 305,623 are objects that we
  have not previously classified as photometric quasar candidates).
  These candidates include 7874 objects targeted as high probability
  potential quasars with $3.5<z<5$ (of which 6779 are new photometric
  candidates).  Our algorithm is more complete to $z>3.5$ than the
  traditional mid-IR selection ``wedges'' and to $2.2<z<3.5$ quasars
  than the SDSS-III/BOSS project.  Number counts and luminosity
  function analysis suggests that the resulting catalog is relatively
  complete to known quasars and is identifying new high-$z$ quasars
at $z>3$.  This catalog paves the way for
  luminosity-dependent clustering investigations of large numbers of
  faint, high-redshift quasars and for further machine learning quasar
  selection using {\em Spitzer} and {\em WISE} data combined with
  other large-area optical imaging surveys.


\end{abstract}


\section{Introduction}


Recent years have seen considerable growth in the number and density
of known actively accreting supermassive black holes in the form of
active galactic nuclei (AGNs) and luminous quasars.  For example,
X-ray surveys now reach AGN densities of more than 9000\,deg$^{-2}$
\citep[e.g.,][]{xlb+11}, albeit over areas of $\ll1$\,deg$^2$.
Spectroscopic follow-up of broad-band optical imaging from the Sloan
Digital Sky Survey-I/II/III (SDSS; \citealt{yaa+00}) project has
expanded the number of confirmed quasars to over 270,000 objects
\citep{srh+10,DR9qsos} over roughly $1/4$ of the sky.  Mid-infrared
(MIR) selection from {\em WISE} and {\em Spitzer} allows AGN selection
(both unobscured and obscured) over the full sky to densities of over
60\,deg$^{-2}$ \citep{sab+12,ask+13}.  Deep large-area optical surveys
such as the Dark Energy Survey (DES; \citealt{des05}) and the Large
Synoptic Survey Telescope (LSST; \citealt{lsst08}) will considerably
expand the number of known AGNs even in already well-mapped areas of
sky, especially at high-$z$ and for low-luminosity AGNs in compact
galaxies.

Our own work has sought to expand the ranks of known quasars by
applying modern statistical techniques to optical imaging data instead
of performing spectroscopy, increasing the number of known quasars to
as many as 1,000,000 \citep{rng+04,rmg+09,bhh+11} and enabling
simultaneous multi-wavelength (optical plus MIR) selection using those
same techniques \citep{rdl+09}.  Such catalogs have enabled
investigations not possible with the density of spectroscopic quasars,
including cosmic magnification \citep{smr+05}, quasar evolution
\citep{mbr+06}, the integrated Sachs-Wolfe Effect \citep{gcn+12},
gravitational lenses \citep{oip+06}, binary quasars \citep{hms+10},
and dust in galaxies \citep{msf+10}---particularly with rigorous
mitigation of the systematics \citep[e.g.,][]{lpm+13,ph13,lp14} that are
inherent to a photometric quasar sample.

The goal of this paper is to extend our previous work as follows: 1)
By providing both optical and MIR data that can be used to help
photometrically identify even larger samples of quasars. 2) Expanding
our pilot optical+MIR quasar selection from $\sim24$\,deg$^2$ in
\citet{rdl+09} to over 10,000\,deg$^2$ by combining optical data from
the SDSS and MIR data from both {\em WISE} and {\em Spitzer}-IRAC.  3)
Using these optical+MIR data to discover new $3.5<z<5.0$
quasars---even in areas that have already received significant
attention (e.g., COSMOS and Bo\"{o}tes).  4) Filling in the gaps of
incomplete redshift from the optically-targeted SDSS-I/II/III
spectroscopic sample.  5) Providing a discovery framework for
clustering studies of high-$z$ quasars within the upcoming {\em
  Spitzer} data within the area of SDSS Stripe 82 as part of the SpIES
project (Timlin, Ross, Richards et al.\ 2015, in preparation).


Section~2 begins with a compilation of over 270,000
spectroscopically confirmed quasars and over 1.5 million
photometrically selected quasars in the SDSS footprint.  These data
are the basis of our training set for further quasar discovery and we
provide this catalog in order to allow others to test their own quasar
selection algorithms and to make meaningful comparison of them to ours
by using the same data set.  In our work, we enhance these data
by matching between the SDSS-optical and the MIR from {\em
  WISE} and {\em Spitzer}, where we have made conversions to put all
of the MIR data on the same photometric system.  Here we emphasize the
difference between our work (which concentrates on finding new type 1
quasars, particularly at high redshift)
and that of \citet{sab+12} and \citet{ask+13} which were designed to
find both type 1 and type 2 AGNs using rigid magnitude and color cuts
to minimize contamination---at the expense of high-redshift quasars \citep{rdl+09,akb+10}.


In Section~3 we describe the construction of our optical+MIR training
sets for distinguishing quasars from stars and apply our selection
algorithm to a test set of objects.  Our primary focus is over
$3.5<z<5.0$ where MIR-only selection is most incomplete
\citep{rdl+09,akb+10,mas+12}; however, we also perform a selection
over $2.2<z<3.5$ and $0<z<2.2$ as our method can also improve upon
optical-only selection which is typically incomplete at $z\sim2.7$ and
$z\sim3.5$ \citep{rsf+06,wp11} and reveals lower-luminosity AGNs at
$z<2.2$ that optical selection alone may fail to distinguish from
compact galaxies.

In Section~\ref{sec:cat} we present our catalog, including photometric
redshifts.  Finally in Section~\ref{sec:analysis} we make comparisons
to previous work, finding that our method allows us to discover many
quasars in hard-to-reach redshift ranges when using either
optical-only or MIR-only selection.  Our $3.5<z<5$ targets are
particularly important for constraining AGN feedback models by
examining the luminosity-dependence of high-redshift quasar clustering
\citep{lhc+06}, where current samples lack sufficient high-redshift
objects over a significant range in luminosity.  We have an
insufficient combination of depth and areal coverage to perform this
analysis with the current sample; however, such analysis can be
performed with {\em Spitzer}-IRAC observations of SDSS ``Stripe 82''
over $\sim110$\,deg$^{-2}$ to a depth of $\sim6\,\mu$Jy (Timlin, Ross,
Richards et al.\ 2015, in preparation).  Section~\ref{sec:analysis}
concludes with a number counts and luminosity function analysis of the
catalog and a discussion of future work.

We report photometry primarily in AB magnitudes, where {\em
  Spitzer}-IRAC Channels 1-2 are given by $[3.6]$ and $[4.5]$, which
are the nominal wavelengths of the bandpasses in microns.  For
comparison with other work using Vega magnitudes we note that the
conversions between {\em Spitzer}-IRAC AB and Vega ($[{\rm
  Vega}]-[AB]$) are 2.788, 3.255, 3.743 and 4.372 mag,
respectively\footnote{http://irsa.ipac.caltech.edu/data/COSMOS/gator\_docs/scosmos\_irac\_colDescriptions.html}.
For example $[3.6]-[4.5] ({\rm Vega}) = [3.6]-[4.5]({\rm AB}) +
0.467$.  For {\em WISE}, we adopt 2.699 and 3.339 as the conversions
to AB from $W_1$ and $W_2$ Vega magnitudes,
respectively\footnote{http://wise2.ipac.caltech.edu/docs/release/allsky/expsup/sec4\_4h.html},
where the {\em WISE} central wavelengths are 3.4, 4.6, 12, and
22\,$\mu$m for $W_1, W_2, W_3$ and $W_4$, respectively.  Cosmology-dependent
parameters are determined assuming $H_o=70\,{\rm
  km\,s^{-1}\,Mpc^{-1}}$, $\Omega_m=0.3$ and $\Omega_{\Lambda}=0.7$,
in general agreement with WMAP results \citep[e.g.,][]{WMAP9}.


\section{The Data}
\label{sec:data}

To conduct our analysis we require optical imaging data of a sample of
objects that require classification; such data will constitute our
{\em test} set.  Some subset of those data must have already been
spectroscopically classified (as quasars) and will form the basis of
our quasar {\em training} set.  These training and test sets will be
described more fully in Section~\ref{sec:train}.  Here we describe the
origin of the data and the parameters determined from the data that
are used for classification by our algorithm.
Section~\ref{sec:master} presents the known quasar sample used to
build the training set, Section~\ref{sec:optical} describes the
optical data, Section~\ref{sec:mir} discusses the infrared data, while
Section~\ref{sec:diags} explores the redshift, magnitude, and color
distributions of the matched optical-infrared data.

\subsection{Master Catalog of Quasars with SDSS Photometry}
\label{sec:master}

In order to optimally select new quasars, we need the largest possible
database of extant quasars with which one can build training sets.  We
construct such a catalog by gathering samples of
spectroscopically-confirmed quasars within the SDSS-I/II/III
\citep{yaa+00,ewa+11} footprint.  Here we detail the input catalogs
and the process used to combine them.  We will refer to this catalog
throughout the paper as the ``master quasar catalog''.

We started with the hand-vetted quasar catalog that concluded the
SDSS-I/II project.  Specifically, Table~5 from \citet{srh+10}, where
we have used the redshifts from \citet{hw10} where available.  The
other large fraction of spectroscopic quasars comes from the the Sloan
Digital Sky Survey-III: Baryon Oscillation Spectroscopic Survey
(SDSS-III/BOSS) project \citep{BOSS}, specifically those quasars
cataloged by \citet{DR10qsos} as part of ``Data Release 10'', where we
used the ``visual inspection'' redshifts.

In addition to the standard BOSS quasars, we include a sample of 851
quasars identified on dates between late 2008 and early 2009 using
Hectospec \citep{ffr+05} on the MMT. The original purpose of this ``MMT"
quasar sample was to investigate the faint end of the quasar luminosity
function in preparation for BOSS, and quasars were targeted using deep
optical data in Stripe 82 and MIR data from Spitzer where available. More
details of these MMT quasars are provided in Appendix C of \citet{rms+12}.
We include {\em all} of these MMT quasars, instead of just those that were
located in Stripe 82, which expands the sample compared to the 444 quasars
listed in Tables 14 and 15 of \citet{rms+12}

Next we add the full quasar catalog from the 2QZ project
\citep{crb+04}\footnote
{www.2dfquasar.org/Spec\_Cat/cat/2QZ\_6QZ\_pubcat.txt}.  The 2dF
instrument provides another catalog input, namely that from the 2SLAQ
project \citep{crs+09}\footnote
{www.2slaq.info/2slaq\_qso/2slaq\_qso\_public.cat} where we have
included only objects labeled as any type of ``QSO''.  The 2dF
instrument has since been upgraded to the AAOmega instrument which was
used to observe objects in our third catalog from the Anglo-Australian
Telescope.  Specifically, we include objects from the AUS project
(Croom et al.\, in preparation), including both a $K$-band limited
sample and a $z>2.8$ selected sample.

We next incorporate quasar data from the AGES project \citep{kec+12},
specifically using data from their Tables~5, 6, and 7.  We have
excluded low-luminosity AGNs by requiring ${\tt qso}=1$ from Table~5.
Quasars from another deep, wide area, namely COSMOS
\citep{COSMOS}\footnote
{irsa.ipac.caltech.edu/data/COSMOS/tables/spectra/} have also been
included in our sample, where the data were limited to type 1 objects \citep{lfr+07,tie+09}.


To increase the number of rare, very high-redshift quasars, we also
include 65 $z>5.8$ quasars from \citet{fsr+06} and \citet{jfa+08}.  The
master quasar catalog was built before a large number of $z\sim5$
quasars were cataloged in Stripe 82 by \citet{mjf+13}, but we recover
49 of the 65 that are bright enough to have matching MIR photometry.


Our master quasar catalog is rounded out by a few smaller samples
of objects meant to extend the range of properties covered.  This
includes the ``BROADLINE'' objects from Table~5 of \citet{pce+06}, the
$z\sim4$ quasars from Table~5 of \citet{gbd+10}, and $KX$-selected
quasars at $z>1$ from \citet[][Tables~4 and 6]{mhp+12}.

There may yet be some known type 1 quasars within the SDSS footprint
that we have not included in our master quasar catalog; however, they should mostly be small
samples of objects that are already represented or much brighter than
the SDSS flux limits (e.g., 3C273 and most ``PG'' quasars from
\citealt{sg83}).

All of the above objects are spectroscopically confirmed quasars;
however, many more likely quasars have been identified
photometrically.  As that information also has value in considering
identification of new quasars, we have included objects listed in the
photometric quasar catalogs of both \citet[][NBCKDE]{rmg+09} and
\citet[][XDQSO]{bhh+11}.

These individual tables are merged together and a flag is set to
indicate the origin.  The flag values run from 0 to 13 as follows,
where spectroscopic redshifts from earlier catalogs in the list trump
later catalogs when there is a duplication: SDSS, 2QZ, 2SLAQ, AUS,
AGES, COSMOS, FAN, BOSS, MMT, NBCKDE, XDQSOZ, PAPOVICH, GLIKMAN,
MADDOX.

For the benefit of those wishing to make use of this master catalog
we make it available in Table~\ref{tab:tab1}.  The
columns are as follows: 1) RA (degrees), 2) Dec (degrees), 3-7) SDSS
run, rerun, camcol, field, and id\footnote{These and other
  SDSS-related information are describe in more detail at
  https://www.sdss3.org/dr9/imaging/imaging\_basics.php.}, 8) the SDSS
morphology ({\tt OBJC\_TYPE}), 9-10) code indicating SDSS data quality
({\tt OBJC\_FLAGS} and {\tt OBJC\_FLAGS}2), 11) SDSS Galactic {\tt EXTINCTION} in all 5
bands, 12) the SDSS flux as measured from Point-Spread-Function
fitting (in nanomaggies) in all 5 bands, 13) the inverse variance of
the PSF flux in all 5 bands, 14) the co-added SDSS PSF flux for those
objects observed in multiple epochs, 15) the inverse variance for
column 14, 16) {\tt PSF\_CLEAN\_NUSE} is an indication of whether there are
multiple epochs of imaging data (values larger than 1 indicate that we
have used the ``CLEAN'' [i.e., co-added] values of the PSF flux in our analysis), 17)
{\tt ZBEST} indicates the redshift determined from each of the sources of
data described in 18) {\tt SOURCEBIT} (numbered 0-13 in the order given
above), 19) indicates whether the SDSS object fell in the ``uniform'' selection
area as described by \citet{rfn+02}, 20-21) codes from the AGES survey
that we used to reject low-redshift AGNs from our training set, 22-25)
photometric redshift information from the NBCKDE photometric quasar
catalog \citep{rmg+09}, 26-28) photometric redshift information from
the XDQSO photometric quasar catalog \citep{bhh+11}.

\begin{deluxetable}{lll}
\tabletypesize{\scriptsize}
\rotate
\tablewidth{0pt}
\tablecaption{Master Quasar Catalog\label{tab:tab1}}
\tablehead{
\colhead{Column} &
\colhead{Name} &
\colhead{Description} \\
}
\startdata
1 & RA & Right Ascension (J2000) \\
2 & DEC & Declination (J2000)  \\
3 & RUN & SDSS run number, see http://classic.sdss.org/dr7/glossary/index.html  \\
4 & RERUN & SDSS rerun number  \\
5 & CAMCOL & SDSS camera column  \\
6 & FIELD & SDSS field number  \\
7 & ID & SDSS ID number (within the field)  \\
8 & OBJC\_TYPE & SDSS object type (stellar$=3$, extended$=6$)  \\
9 & OBJC\_FLAGS & SDSS object flags, see http://classic.sdss.org/dr7/products/catalogs/flags.html  \\
10 & OBJC\_FLAGS2 & SDSS object flags  \\
11 & EXTINCTION & Magnitudes of Galactic extinction in ugriz  \\
12 & PSFFLUX & Point-spread-function flux in ugriz  \\
13 & PSFFLUX\_IVAR & Inverse variance of point-spread-function flux in ugriz  \\
14 & PSFFLUX\_CLEAN & Co-added point-spread-function flux in ugriz  \\
15 & PSFFLUX\_CLEAN\_IVAR & Inverse variance of co-added point-spread-function flux in ugriz  \\
16 & PSF\_CLEAN\_NUSE & Flag indicating whether co-added (CLEAN) flux should be used  \\
17 & ZBEST & Spectroscopic and photometric redshifts from the sources indicated by SOURCEBIT  \\
\multirow{2}{*}{18} & \multirow{2}{*}{SOURCEBIT} & Bitwise flag from $2^0$ to $2^{13}$ indicating the redshift source as coming from SDSS, 2QZ, AUS, AGES,\\
 & & COSMOS, FAN, BOSS, MMT, NBCKDE, XDQSOZ, PAPOVICH, GLIKMAN, MADDOX, respectively \\
19 & SDSS\_UNIFORM & Indicates whether the SDSS object fell in the ``uniform'' selection area, see Richards et al. (2002)  \\
20 & AGES\_QSO & AGES flag, see Kochanek et al. (2012)  \\
21 & AGES\_CODE06 & AGES flag, see Kochanek et al. (2012)  \\
22 & KDE\_ZPHOTLO & Minimum photometric redshift from Richards et al. (2009)  \\
23 & KDE\_ZPHOTHI & Maximum photometric redshift from Richards et al. (2009)  \\
24 & KDE\_ZPHOTPROB & Photometric redshift probability from Richards et al. (2009)  \\
25 & KDE\_LOWZORUVX & Flag indicating a UV-excess or low-redshift source; Richards et al. (2009)  \\
26 & XDQSOZ\_PEAKPROB & Peak of the redshift probability from Bovy et al. (2011)  \\
27 & XDQSOZ\_PEAKFWHM & FWHM of the redshift peak from Bovy et al. (2011)  \\
28 & XDQSOZ\_NPEAKS & Number of peaks in the Bovy et al. (2011) photo-z distribution
\enddata
\end{deluxetable}

\subsection{Optical Data}
\label{sec:optical}

Over more than 10 years, the SDSS used a sophisticated telescope
\citep{gsm+06} fitted with a large field-of-view camera \citep{gcr+98}
to take exposures through $ugriz$ filters \citep{fig+96}. For the
training and testing sets in this paper, we use the ``Data Release 9''
(DR9) versions of this SDSS imaging \citep{DR9}.  DR9 included the
latest astrometric and photometric calibrations for imaging in the
original northern SDSS footprint and in the southern footprint that
was added as part of SDSS DR8 \citep{DR8}. Specifically, we use the
versions of the SDSS imaging provided in the {\tt calibObj} or ``data
sweep''
files\footnote{http://data.sdss3.org/datamodel/files/PHOTO\_SWEEP/RERUN/calibObj.html}
that are discussed in \citet{bss+05}. We limit the data sweeps to only
objects that are {\tt PRIMARY} in SDSS imaging (e.g., see Table 5 of
\citealt{slb+02}), but do {\em not} further restrict our optical
sources using cuts on image quality flags at this stage (any
additional flag cuts are described in the relevant sections of this
paper). We use such {\tt PRIMARY} sources from the SDSS data sweep
files as our test data and also match our heterogeneous master
training catalog of quasars (described in the previous section) to
{\tt PRIMARY} objects from these data sweeps.

While the spectroscopic identifications that we tabulate have a
heterogeneous origin, one advantage of the catalog of quasars that we
have built is that their optical imaging is derived solely from the
SDSS imaging camera \citep{gcr+98}, providing a homogeneous aspect to
the data set.

All of the optical magnitudes are reported in the catalog are asinh
PSF magnitudes \citep{lgs+99} corrected for dust extinction using the
coefficients given by \citet{sf11}.  Fluxes are reported in
nanomaggies without any dust extinction correction.  The full list of
cataloged parameters are given in Table~\ref{tab:tab1} for the master
quasar catalog and Section~\ref{sec:cat} for our quasar candidate
catalog; further information on each source is publicly available.



\subsection{Infrared Data}
\label{sec:mir}

To create our MIR data set, we begin by merging large areas of
relatively deep {\em Spitzer}-IRAC data \citep{fha+04} with shallower,
but wider-area {\em WISE} data \citep{WISE}.  This has the advantage
of allowing us to probe both a wide area and relatively deep (in a
fraction of that area).

The {\em WISE} data come from the ALLWISE data
release\footnote{http://wise2.ipac.caltech.edu/docs/release/allwise/},
where we have kept only objects with {\em both} $W_1$ and $W_2$ detections
and have excluded objects that do not meet the following quality
control criteria: {\tt w1flg $<=1$ \&\& w2flg$<=1$} (to avoid sources
with bad pixels or that are upper limits), {\tt cc\_flags=='0000'} (to
avoid objects affected by diffraction spikes, ghosts, latent images,
and scattered light), {\tt ext\_flg==0} (to limit to MIR point
sources), and {\tt w1snr$>2$ \&\& w2snr$>2$} (to limit to objects that
are well-detected in both $W_1$ and $W_2$)\footnote{See
  http://wise2.ipac.caltech.edu/docs/release/allwise/expsup/sec2\_1a.html
  for detailed explanation of these parameters.}.  By matching known
SDSS quasars to ALLWISE, we estimate that these cuts cull 9.6\%,
3.0\%, 0.6\%, and 0.2\% of real sources, respectively.  This
incompleteness is corrected in our number counts and luminosity
function analysis in Section~\ref{sec:analysis}.


The {\em Spitzer} catalogs include 1) the SWIRE data \citep{lsr+03},
2) the XFLS data \citep{lacy05}, 3) the COSMOS data \citep{scosmos07},
4) our own pilot sample of {\em Spitzer}-IRAC data centered on known
high-$z$ quasars in SDSS Stripe 82 (data tabulated in
\citealt{krm+13}), 5) the SDWFS data in the Bo\"{o}tes field
\citep{esb+04}, and 6) the SERVS data \citep{SERVS}.  The SWIRE, XFLS,
COSMOS, and SDWFS data are the same data used in \citet{rdl+09}; see
that paper for more details.  Bo\"{o}tes data are taken from
\citet{SDWFS}, specifically {\tt SDWFS\_ch1\_stack.v34.txt}, adopting
the aperture-corrected 4$\arcsec$ (diameter) flux densities.  This
catalog corresponds to a depth of 12$\times$30s and we have limited to
objects detected in both Channels 1 and 2 and with SExtractor flags of
0 or 2.  Vega magnitudes have been converted to $\mu$Jy.
The SERVS data are described in detail in \citet{SERVS}.

Our Stripe 82 data includes pointed observations of over 300 known
$z>2$ quasars in the SDSS Stripe 82 field \citep{Annis11,jfb+14} and were
processed in a manner similar to that which was used for the SWIRE
data set.  Photometry for these sources is tabulated in
\citet{krm+13}.  We report fluxes in a $1\farcs9$ aperture radius.

For all of the above data sets, we have included all objects that are
not flagged by SExtractor \citep{SExtractor} as blended in either IRAC
Channel 1 or Channel 2 and we have applied no explicit flux limits to
the individual catalogs.  Flux densities have been converted to
$\mu$Jy if the original data have other units.  We report errors that
have been increased by 3\% (10\% for XFLS) in quadrature since
SExtractor only reports the RMS at the image position; this is
consistent with \citet[][Section~4]{dkb+12}.

We would like to be able to use MIR measurements from both {\em
  WISE} and {\em Spitzer}; however, photometry from these two
spacecraft are on different photometric systems.  There is strong
similarity in the two shortest wavelength filters of the systems, but
a correction needs to be applied.  As such, the {\em WISE} data have
been converted from Vega magnitudes on the {\em WISE} system to
$\mu$Jy in the {\em Spitzer}-IRAC system using color terms appropriate
for each of the individual objects (based on their $W_1-W_2$ colors).
This process is important for allowing us to treat the {\em WISE} and
{\em Spitzer} data equivalently.  As the $W_3$ and $W_4$ data are much
shallower than $W_1$ and $W_2$, we have only tabulated the $W_1$ and $W_2$
photometry and we have only kept objects with detections in both of
those bands.

As an illustration of our conversion of the {\em WISE} Vega system to
{\em Spitzer} AB, we convert the $W_1-W_2 \; ({\rm Vega})=0.8$ color-cut
used by \citet{sab+12} to the {\em Spitzer} AB system.  First we find
that
\begin{equation}
W_1({\rm Vega}) - W_2({\rm Vega}) = (W_1({\rm AB}) - 2.699) - (W_2({\rm AB}) - 3.339)
\end{equation}
so that the above cut is $W_1-W_2 ({\rm AB}) = W_1-W_2 ({\rm Vega}) -
0.64 = 0.16$.  We have then created a look-up table for the conversion
of {\em WISE} AB magnitudes to {\em Spitzer} AB magnitudes as a
function of color (assuming a power-law SED).  In general these
corrections are small for $W_1$ and $W_2$; see \citet[][Table1]{WISE}.
We find that at $W_1-W_2 ({\rm AB}) = 0.16$: $[3.6] = W_1 ({\rm AB}) -
0.028$ and $[4.5] = W_2 ({\rm AB}) + 0.013$, so that $W_1-W_2 ({\rm
  Vega}) = 0.8$ is equivalent to $[3.6]-[4.5] ({\rm AB}) = 0.119$.
Similarly we can convert a $W_2 ({\rm Vega}) = 15.05$ magnitude cut
(at this color) to {\em Spitzer} AB as follows: $[4.5] ({\rm AB}) =
W_2 ({\rm Vega}) +3.339 +0.013 = 18.402$.  We illustrate these cuts in
Section~\ref{sec:train}, where for the sake of simplicity we have
ignored the color-dependence of the magnitude limit.  As the agreement
with {\em Spitzer} photometry has significantly improved for the
ALLWISE data release as compared to the older, All-Sky {\em WISE}
data, we have not further corrected for the remaining offsets.  The
typical ALLWISE limits are 54\,$\mu$Jy in $W_1$ or 16.9 in Vega mags
and 71\,$\mu$Jy in $W_2$ or 15.9 in Vega mags, but depend on location
due to {\em WISE}'s polar orbit.  In AB mags, these limits are 19.6
and 19.3.  See the ALLWISE Explanatory
Supplement\footnote{http://wise2.ipac.caltech.edu/docs/release/allwise/expsup/sec2\_3a.html}
for a discussion of how the {\em Spitzer} and {\em WISE} differences,
$[3.6]-W_1$ and $[4.5]-W_2$, behave as a function of magnitude and
for information on how the {\em WISE} sensitivity changes with
coordinate.



We generate a single merged MIR catalog by matching the above data sets
using a 2$\arcsec$ matching radius with priority being given to
objects from the individual catalogs as follows: SERVS, SWIRE, COSMOS,
SDWFS, XFLS, Stripe82, and {\em WISE}.  That is a SERVS detection will
overwrite a SWIRE detection.  Only one {\em Spitzer} detection of
each object was allowed and a flag 
was set to indicate which catalog the photometry comes from.
However, if there is data from both {\em WISE} and {\em Spitzer}, we
have also kept the {\em WISE} data for reference.

This final MIR catalog is then matched to the SDSS-III imaging data
using a 2$\arcsec$ matching radius.  No explicit flux limits have
been applied.  Dust extinction has been corrected as
$A_{[3.6]}=0.197E(B-V)$ and $A_{[4.5]}=0.180E(B-V)$, consistent with
\citet{ccm89} as reported by the NASA/IPAC Infrared Science
Archive\footnote{irsa.ipac.caltech.edu}.

The full SDSS-III footprint lacks deep near-IR imaging, since 2MASS
\citep{sss+97} is too faint to provide counterparts for the bulk of
our quasar sample.  However, when available, near-IR data is very
useful for improving photometric redshift (photo-$z$) estimates.
Thus, while we do not use near-IR data for our quasar selection
algorithm, we do match our optical catalog to near-IR catalogs from
the regions of sky covered by
the UKIRT Infrared Deep Sky Survey (UKIDSS; \citealt{UKIDSS}) and the
Vista Hemisphere Survey (VHS; \citealt{VHS}).  We used a matching
radius of 1$\arcsec$ and included only objects that have measurements
in each of $J$, $H$, and $K$.  While these near-IR data are not
simultaneous with the optical or MIR data, which causes some scatter
in the color distributions, even simultaneous observed-frame
multi-wavelength (and thus multi-distance scale) data would not be
simultaneous in the rest-frame.

\begin{figure}[h!]
\epsscale{0.9}
\plotone{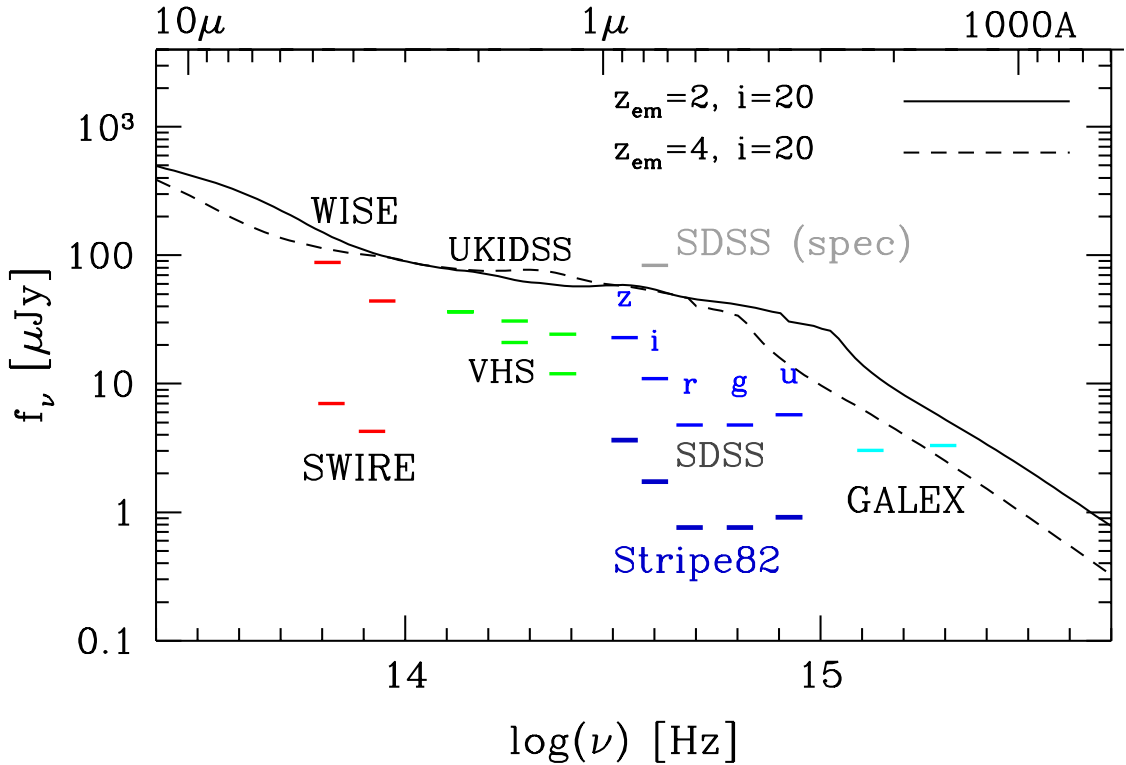} 
\caption{Relative limits of the multi-wavelength data.  The bars indicate the effective wavelength of the bandpasses, but are not scaled to represent the size of the bandpass.  Red indicates MIR data from ALLWISE and {\em Spitzer}-SWIRE, green indicates the limits of UKIDSS and VHS, blue shows the depth of both single-epoch and multi-epoch (Stripe 82) SDSS photometry, while cyan gives the limits of the {\em GALEX} AIS survey.  Two example quasars spectral energy distributions (from \citealt{krm+13}) are given for $z=2$ (solid black line) and $z=4$ (dashed black line), both corrected for Lyman series extinction and normalized to $i=20$, which is roughly the limit of SDSS spectroscopy for high-redshift (it is $i=19.1$ for low redshift, which is shown in gray).
 \label{fig:fig1}}
\end{figure}

Figure~\ref{fig:fig1} shows the relative limits of the MIR and near-IR
data as compared to the optical for a typical quasar spectral energy
distribution \citep{krm+13}.  High-$z$ quasars found from SDSS
photometry with $i<20$ are expected to be detected in ALLWISE.  They
should also be detected by UKIDSS and would be detected by {\em GALEX}
in the bluest bandpass.  Quasars closer to the SDSS photometric limit
(for single-epoch data) can be much fainter than the ALLWISE, UKIDSS,
and VHS limits, which will limit the completeness of this catalog.
Fainter quasar candidates are limited by the depth of ALLWISE (or the
area of {\em Spitzer}).

\subsection{Diagnostics}
\label{sec:diags}

\begin{figure}[h!]
\epsscale{0.8}
\plotone{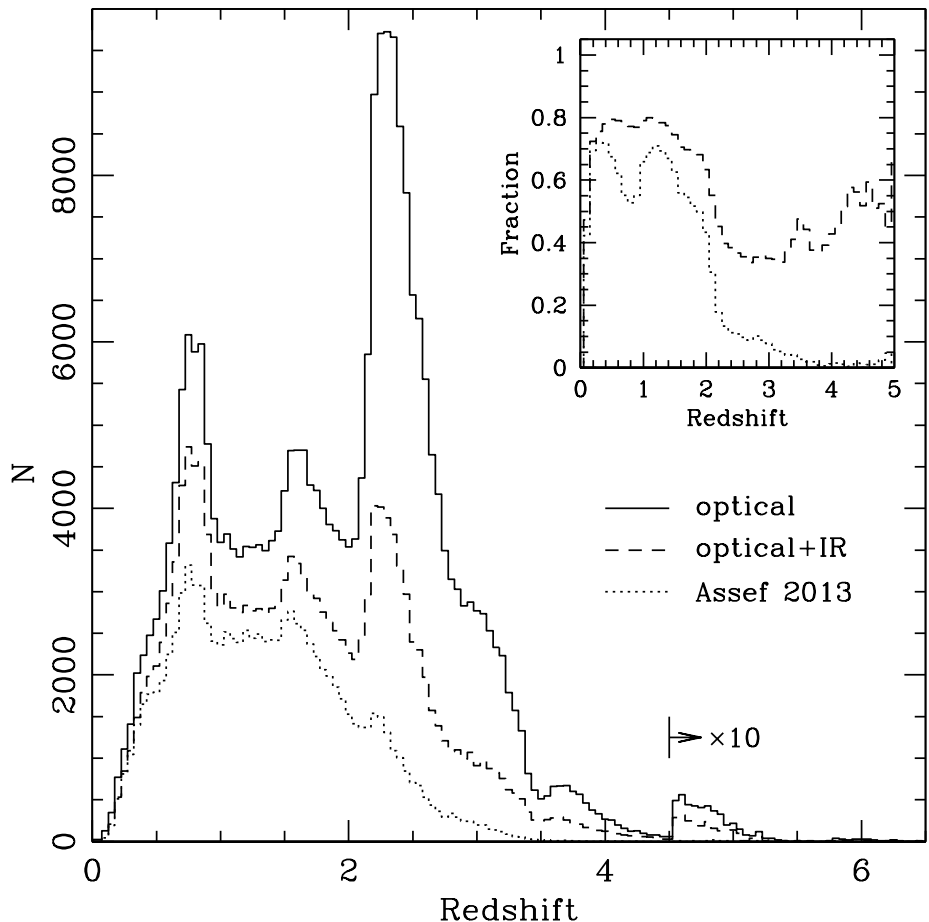} 
\caption{Redshift distribution of the full spectroscopic quasar sample ({\em solid line}; 274,329 quasars), for the IR-matched sample ({\em dashed line}; 157,701 quasars---the parent sample of our quasar training sets), and for the IR-matched sample with the 75\% reliability limit from \citet{ask+13} imposed ({\em dotted line}).    Beyond redshift 4.5 the distributions have been scaled by a factor of 10 to better show the high-$z$ part of the samples.  The inset gives the ratio of the dashed line to the solid line and the dotted line to the solid line.  Losses at low redshift are dominated by flag cuts ($\sim13\%$, independent of redshift).  Further losses at high redshift are primarily due to implicit ({\em dashed line}) or explicit ({\em dotted line}) magnitude limits of the sub-samples as can be seen in Figure~\ref{fig:fig3}.
 \label{fig:fig2}}
\end{figure}

Here we provide some diagnostic plots to illustrate the range of
optical and MIR properties spanned by our choice of data.
Figure~\ref{fig:fig2} shows the redshift distribution for all of the
objects in our master quasar catalog, including those objects where
only optical photometry is available and those objects where MIR
photometry exists.  The peaks in redshift in this figure represent
selection effects.  The SDSS DR7 quasar sample peaked at $z\sim1.5$,
while the SDSS DR10 quasar selection was optimized for $z\sim2.5$,
with contamination coming at $z\sim0.8$.  Most of the losses of
IR-matched objects at low redshift are due to the flag cuts imposed
upon the {\em WISE} data.  At high redshift, the difference between
the focus of our work (not relying on MIR color cuts) and that of
\citet{ask+13} (which utilizes MIR color cuts) is readily apparent.

\begin{figure}[h!]
\epsscale{0.8}
\plotone{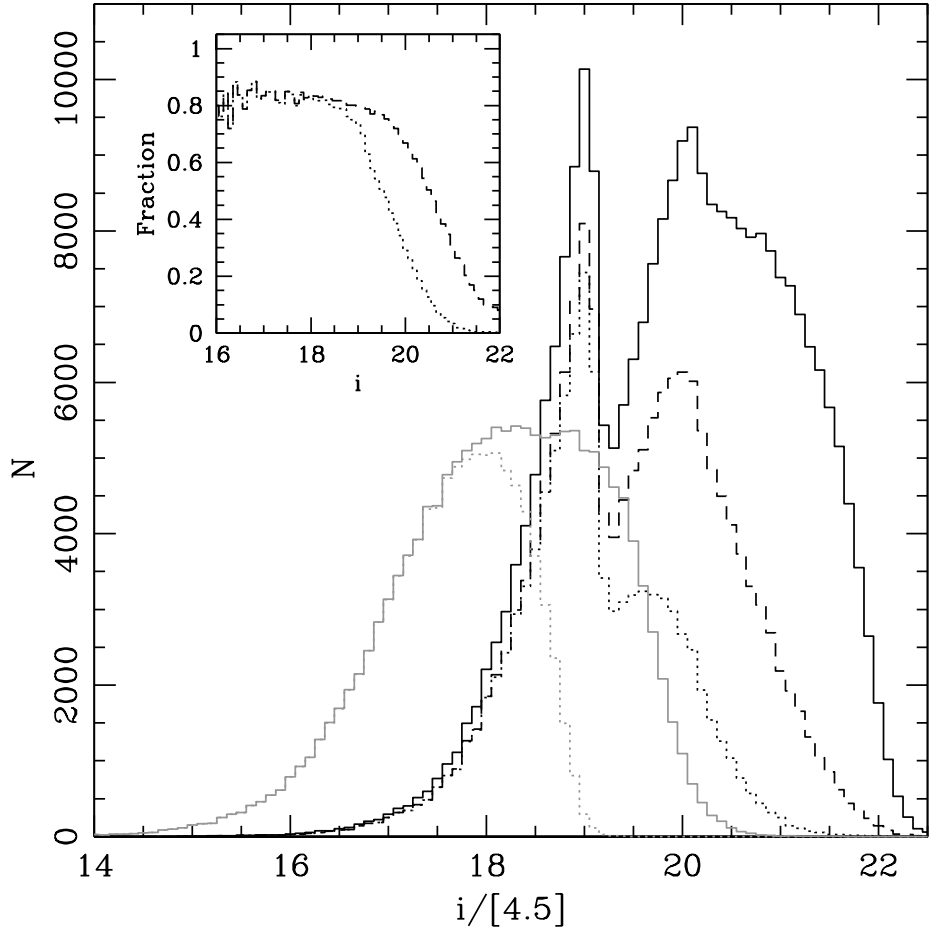} 
\caption{$i$-band magnitude distribution of the full spectroscopic quasar sample ({\em solid black line}), for the IR-matched sample ({\em dashed black line}---the parent sample of our quasar training sets) and for the IR-matched sample with the 75\% reliability limit from \citet{ask+13} imposed ({\em dotted black line}).    The inset shows the ratio of the latter two samples to the full sample, demonstrating that our matching to {\em WISE} (and/or {\em Spitzer}) photometry is over 80\% complete to $i\sim19$ ({\em dashed line}) and that our greater sensitivity to high-redshift quasars relative to \citet[{\em dotted line}][]{ask+13} is largely due to probing deeper.  The gray histograms in the main panel show the distribution in [4.5] for our full training set ({\em solid}) and after imposing the 75\% reliability cut of \citet{ask+13} ({\em dotted}).  
 \label{fig:fig3}}
\end{figure}

Figure~\ref{fig:fig3} shows the magnitude distribution of the objects
in the master catalog.  The peaks in the distribution are caused by a
combination of magnitude limits: the SDSS DR7 quasar sample had a
$z<3$ magnitude limit of $i<19.1$ and a $z>3$ limit of $i<20.2$, while
SDSS DR10 probed to $g<21.85$ ($i\sim22$).  Although adding MIR
photometry is very powerful for AGN selection, it is also responsible
for reducing the completeness to known quasars by a factor of $\sim$2
by $i=20$.  Up to $i\sim19$, over 80\% of our quasar sample includes
IR measurements from {\em WISE} or {\em Spitzer}.  
Most of the losses at bright magnitudes occur due to our attempts to
restrict ourselves to the highest quality {\em WISE} data as noted
above.  The fraction of bright quasars with IR matches is roughly
consistent with the expected loss of $\sim13$\% of sources due to the
flag cuts on the {\em WISE} data and the fraction found by
\citet{whj+12}.  
That is, the curves in the insets of Figures~\ref{fig:fig2} and
~\ref{fig:fig3} should be shifted up by 0.13 to correct for objects
removed due to flag cuts.  The dotted lines show the effect of the
\citet{ask+13} reliability cuts relative to the objects in our
training set (dashed lines).

\begin{figure}[h!]
\plotone{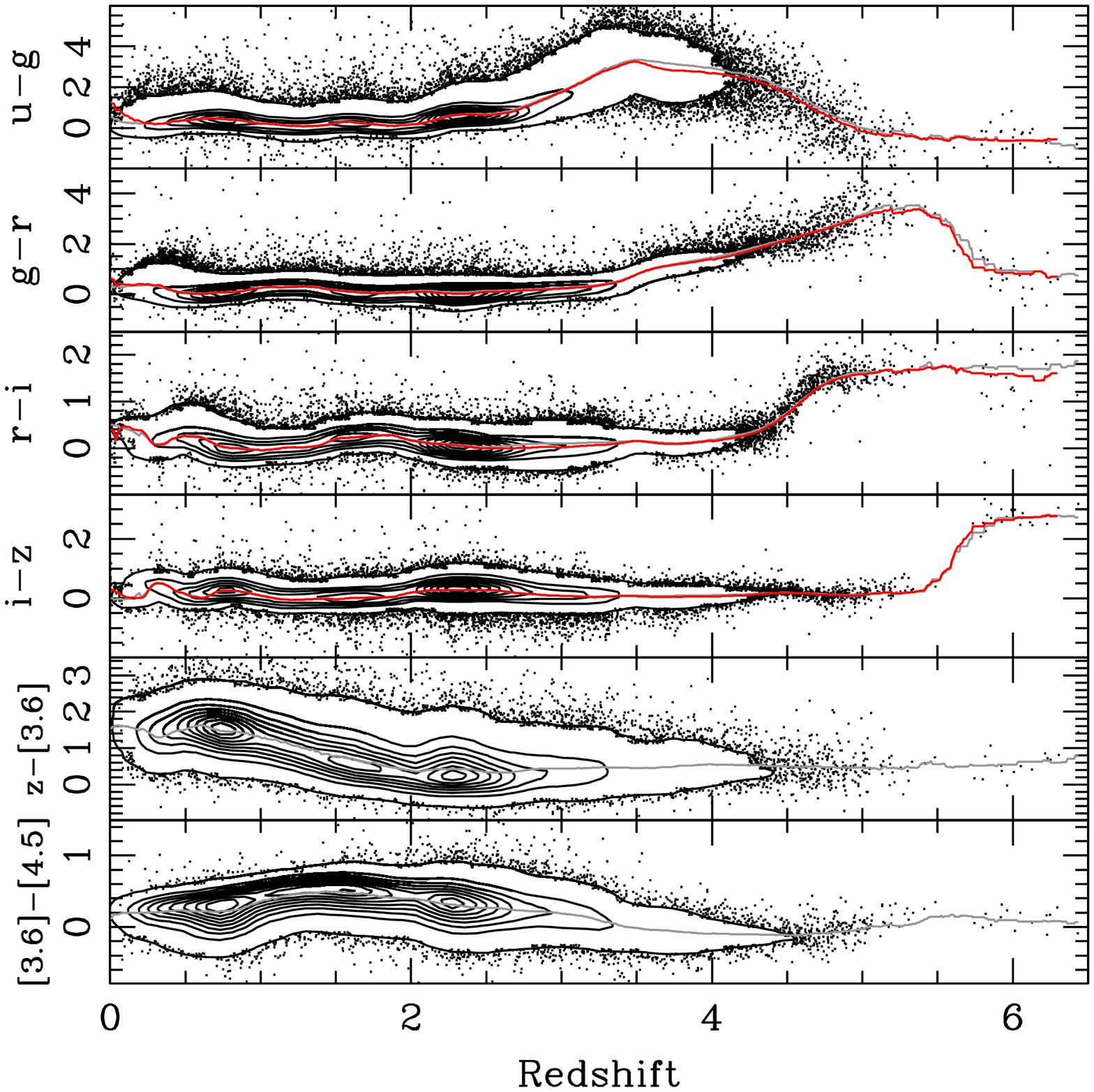} 
\caption{Color vs.\ redshift for the spectroscopic quasar sample.  Black (linear) contours and dots show the color distributions of the individual objects.  The top four panels include all of the spectroscopic objects; the bottom two panels contain only those matching to the IR sample.  Lines give the mean color-redshift relations (which are used to compute the photometric redshifts).  The red line is for all of the optical data, while the gray line shows the mean for the objects that additionally have IR matches.  In the top four panels there is good agreement between the red and gray lines (and thus between the quasars with and without matching IR photometry).
\label{fig:fig4}}
\end{figure}

Figure~\ref{fig:fig4} shows the quasar colors as a function of
redshift.  In addition to the data points, we also depict the mean
colors as a function of redshift for both the full optical sample and
the more limited optical+MIR sample.  Overall, there is good agreement
between the samples.

\section{Classification}

In Section~\ref{sec:data} we tabulated quasars both with and without
MIR photometry; for the remainder of this paper we will consider only the
optical+MIR data set.  After building training and test sets
(Section~\ref{sec:train}) in a similar manner to that described in
\citet{rdl+09}, we will apply the same Bayesian selection algorithm
(Section~\ref{sec:algorithm}) described in our previous papers, and
then we will describe the selection results (Section~\ref{sec:class}).

\subsection{Training and Test Sets}
\label{sec:train}


Starting with the matched optical+MIR photometry (both for known
quasars and all SDSS-DR10 sources), we create the test set (objects to
be classified) along with the quasar and non-quasar (``star'') training sets
as follows.

We first restrict the data to objects that are expected to have ``clean''
photometry, which, for our purposes, we define based on whether or not
they have any of the following SDSS imaging quality flags set: {\tt
  INTERP\_PROBLEMS}, {\tt DEBLEND\_PROBLEMS}, {\tt NOT\_BINNED1}, {\tt
  EDGE}, {\tt BRIGHT}, {\tt SATUR}, {\tt MOVED}, {\tt BLENDED}, {\tt
  NODEBLEND}, and {\tt NOPROFILE}. These flags are fully defined in
Table 9 of \citet{slb+02} except for {\tt INTERP\_PROBLEMS}, {\tt
  DEBLEND\_PROBLEMS} and {\tt MOVED} which are detailed in
\citet{rfn+02} and/or are further discussed in Appendix A of
\citet{rms+12}.
Objects must also have flux values of $<1000$ nanomaggies ($m_{\rm
  AB}>15$) in all bands to be included as brighter fluxes can lead to
saturated pixels.
However, we have made this cut before applying any dust extinction
corrections, so objects that are intrinsically brighter than $m_{\rm
  AB}=15$, but that are not saturated in the images are kept.  

If good {\em co-added} (multi-epoch) photometry is reported in all
bands (as indicated by {\tt PSF\_CLEAN\_NUSE}$>0$)\footnote{Again see
  http://data.sdss3.org/datamodel/files/PHOTO\_SWEEP/RERUN/calibObj.html
  for descriptions of the format of the data sweeps files that we
  use.}, then we retain the co-added fluxes (and errors); otherwise
the single-epoch fluxes are used.  To handle the problem of negative
fluxes we have used the asinh magnitude prescription of
\citet{lgs+99}.

Initially our classification included both point and extended
(optical) sources as have our previous catalogs. Later we will
restrict our analysis to just the point sources.  
At this point, the test set consists of all the photometry from all of the
``good'' point and extended sources described above.  No further
restrictions are placed on the objects that we attempt to classify.
The classification parameters are the set of adjacent colors determined
from each of the 5 optical and 2 mid-IR magnitudes that we catalog,
specifically: $u-g$, $g-r$, $r-i$, $i-z$, $z-[3.6]$, and
$[3.6]-[4.5]$.  In all there were 50,225,630 objects in the test set.

The quasar training set is the subset of the test set for which there
is a match in the master quasar catalog with a spectroscopic redshift
(i.e., we have not included photometric quasars) as noted in Section~\ref{sec:master}.
The ``stars'' training set is again a subset of the test set.  Here
sources matched to known (spectroscopic) quasars are excluded.  The
final stars training set is a randomly selected sample of
$\sim$700,000 objects (taking those objects where the hundredths and
thousandths digits of the IRAC CH2 flux density were ``01'').  The
vast majority of these objects lack spectroscopic classification as
stars, thus these are not only stars, but can be (compact) galaxies (and
previously unidentified quasars); see the discussion of the cleaning
process below.  Thus ``stars'' in this context is shorthand meaning optical
point sources that have not been classified as quasars in the redshift
range we are trying to select.


In practice we have actually made three pairs of quasar and star
training sets as quasar colors change considerably at high redshift
and it is best to treat them as separate populations.  Thus, the
quasar training sets are created by parsing through the quasars and
keeping only those within the redshift range of interest.  Quasars
outside of that redshift range are put into the ``stars" training set.
The three ranges used are $0<z<2.25$ (11984 quasars), $2.15<z<3.55$
(45561 quasars), and $3.45<z<5.5$ (3321 quasars), where the overlap is
to minimize the loss of objects near the redshift boundaries and we
stop at $z=5.5$ since selecting higher redshifts generally requires
additional care \citep{fsr+06}.  We will refer to objects selected
from the use of training sets focusing on these redshift ranges as
``low-$z$'', ``mid-$z$'', and ``high-$z$'' throughout the rest of the
paper.

\begin{figure}[h!]
\plotone{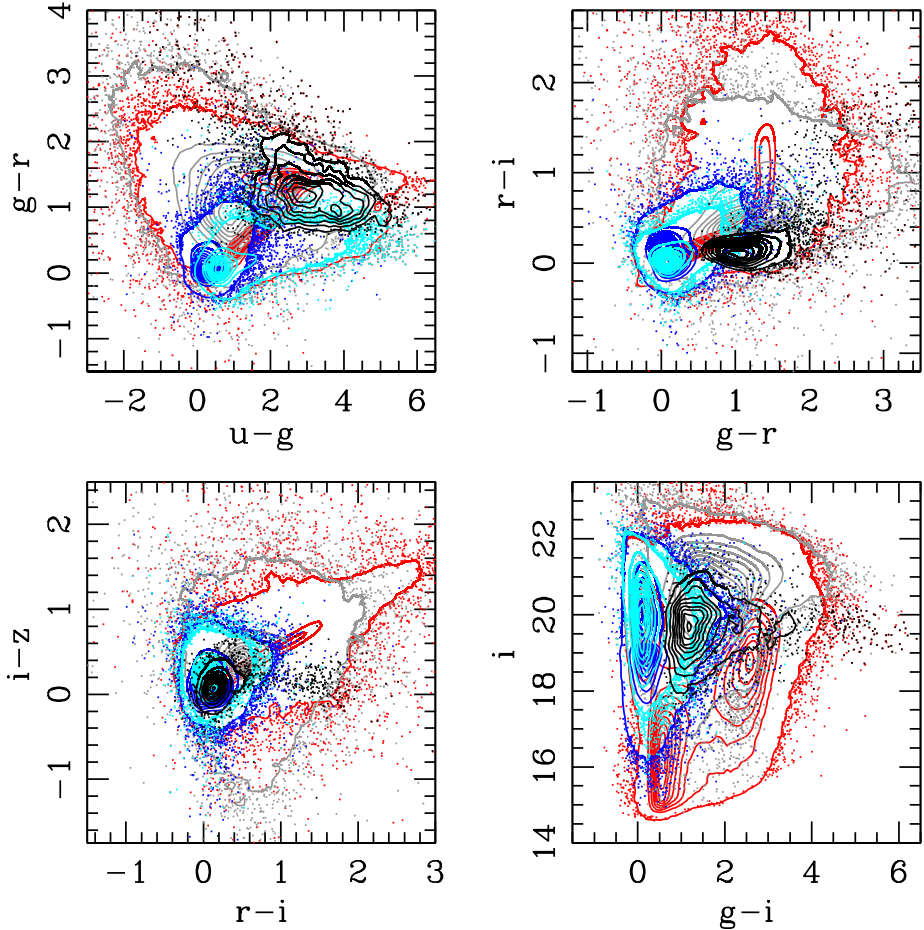}
\caption{Optical colors of training set objects.  Point sources are in red, extended sources in gray, high-$z$ quasars in black, mid-$z$ quasars in cyan, and low-$z$ quasars in blue.  Extended sources are not explicitly included in the training set but are shown here for reference given that separation of point and extended sources is not perfect (particularly at faint magnitudes).
\label{fig:fig5}}
\end{figure}

\begin{figure}[h!]
\plotone{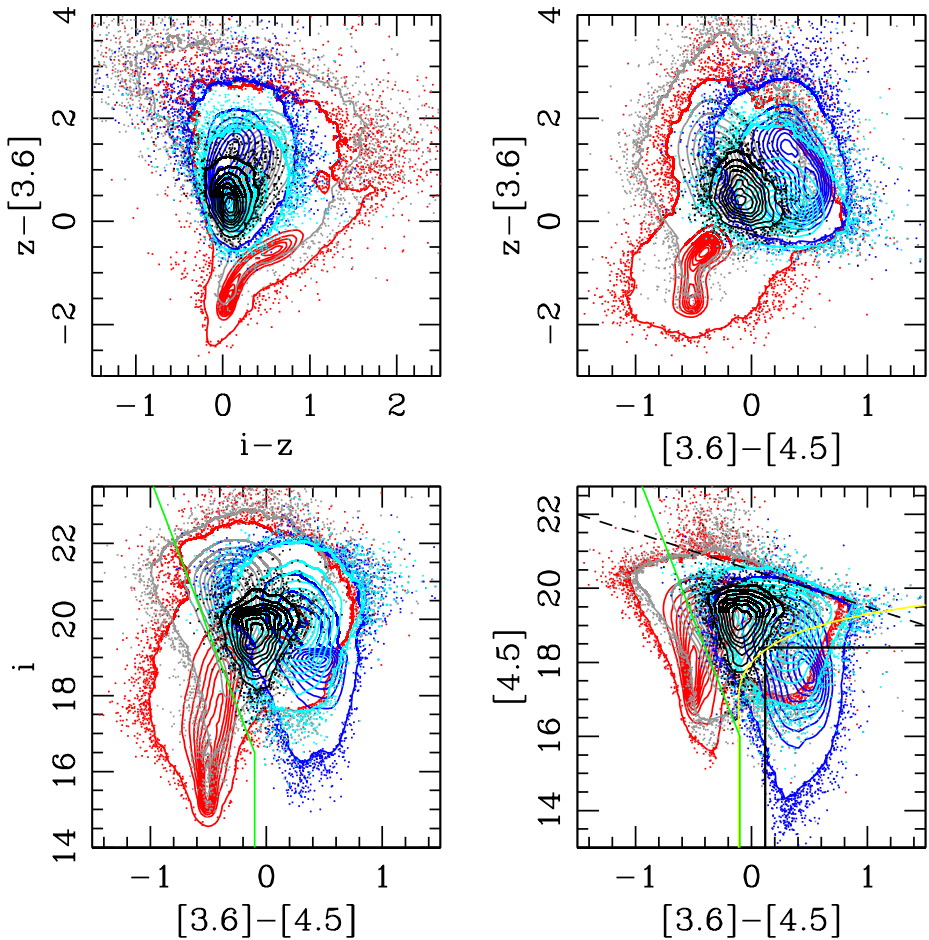}
\caption{MIR colors of training set objects.  Point sources are in red, extended sources in gray, high-$z$ quasars in black, mid-$z$ quasars in cyan, and low-$z$ quasars in blue. The dashed black line shows the detection limit as a function of color for a theoretical object with [3.6]=20.5.
The solid black lines indicate the color and magnitude limits of the \citet{sab+12} selection in AB magnitude space, while the yellow curve gives the 75\% reliability selection from \citet{ask+13}.  The green lines in the bottom panels give our own cuts that are intended to reduce stellar contamination; these are not applied to the training sets, but are shown here for comparison to the test set output.
  \label{fig:fig6}}
\end{figure}

Figure~\ref{fig:fig5} presents the optical colors (and a magnitude) of
the objects in our training sets.  For the star training set, we show
only the low-$z$ training set which includes quasars above $z=2.2$.
All three quasar training sets are shown.  Similarly,
Figure~\ref{fig:fig6} gives the MIR colors of the training set
objects.  Here we include the color-magnitude cuts ({\em solid black
  line}) used by \citet{sab+12} to select their quasar sample in
addition to the (somewhat more inclusive) 75\% reliability selection
({\em solid yellow curve}) of \citet{ask+13}.  Comparison of these
curves to the distribution of high-redshift quasars illustrates their
bias against such objects as shown in Section~\ref{sec:diags}.  This
reflects a conscious choice to be sensitive to both type 1 and type 2
AGNs without significant contamination from inactive galaxies.  Our
approach is complementary in that we will endeavor to be as complete
as possible to high-redshift type 1 quasars, at the expense of type 2
quasars.  The green lines in Figure~\ref{fig:fig6} depict the cuts
that we will use to reduce stellar contamination from the test sets as
shown in Section~\ref{sec:class}.  We duplicate them here to emphasize
that they would throw out relatively few of the training set objects.

\subsection{Application of the Algorithm}
\label{sec:algorithm}

\begin{figure}[h!]
\epsscale{0.8}
\plotone{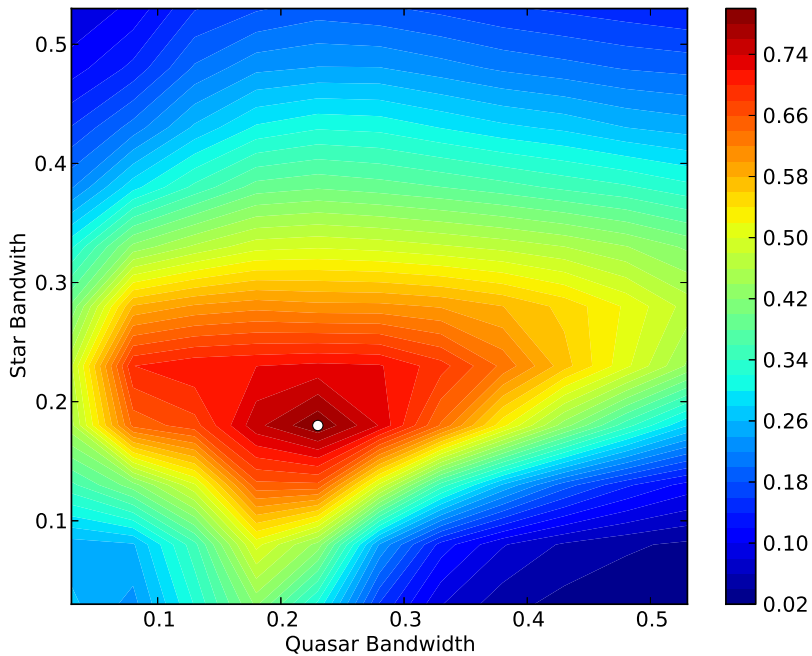}
\caption{Graphical depiction of the search for optimal bandwidths for the star and quasar training sets.  The color bar indicates the ``rating'' of each bandwidth pair, which is determined by the product of the self completeness and efficiency.  The optimal bandwidth in this case (the high-$z$ training set) was found to be (0.23,0.18) for (quasars,stars).
  \label{fig:fig7}}
\end{figure}

As described in more detail in \citet{rng+04,rmg+09,rdl+09}, our
algorithm requires that we compute a ``bandwidth'' that best describes
the range of colors of each object class.  This is akin to determining
the best bin size to represent one's data in a histogram: too many
bins leaves too few objects in each bin, while too few bins
over-smooths the data and causes a loss of information.  Thus, the
bandwidth is essentially a smoothing parameter for the color
distributions.  These bandwidths are determined by a
self-classification step, choosing the bandwidth that yields the most
complete recovery of known quasars with the smallest contamination
from stars.  As in our previous work, we first perform an initial
self-classification of the training sets, then we throw out objects
initially classified as quasars from the star training set (since we
expect our star sample to be contaminated by those very objects that
we wish to recover where other algorithms have failed).  The final
bandwidth is determined from the original quasar training set and the
``cleaned'' star training set.  An example ``heat map'' showing the
minimization of the bandwidths for self-classification of stars and
quasars in the high-$z$ training sets is shown in
Figure~\ref{fig:fig7}.  Optimal bandwidths were computed for each
of the low-$z$, mid-$z$, and high-$z$ training sets.

The only other free parameter in our classification is the Bayesian
stellar prior, which represents our expectation of what fraction of
objects are really stars.  For low-$z$ classification this was set to
98\% (that is we expect 98\% of the objects in the test set to be
``stars'').  For the mid-$z$ classification it was set to 99.9\%,
reflecting the lower density of quasars in this redshift range as
compared to lower redshift.  Finally for high-$z$ classification, it
was set to 99.99\%.  These numbers are estimated from the ratio of the
number of objects in the test set to the number of objects in the
training set, which provides a conservative estimate of the quasar
fraction.  These star priors demonstrate the level to which quasar
classification is a ``needle in a haystack'' problem that requires
methods more sophisticated than simple color cuts.  Note that small
changes in the prior only make small changes in the number of quasars
selected.  For example, in the low-$z$ case, lowering the stellar
prior by 1\% does not increase the number of quasar candidates by 1\%
of the test set (roughly a half million objects); rather we find that
it changes the number by roughly 1\% of the quasar candidates ($\sim
7000$ objects).

\subsection{Classification Results}
\label{sec:class}

Here we present the results of our classification.  This process is an
extension of the 8-D (optical plus MIR colors) selection described in
\citet{rdl+09}, using the algorithm described in more detail by
\citet{rng+04,rmg+09}.  

Our algorithm can roughly be summarized as choosing objects for which 
\begin{equation}
P({\rm colors}|{\rm quasar})P({\rm quasar})>P({\rm colors}|{\rm star})P({\rm star}),
\end{equation}
where P(star) is the stellar prior, P(quasar) is $1-$P(star) and
P(colors$|$quasar) is the probability of an object having certain
colors given that it is known to be a quasar (and similarly for the
stars).  In practice we have performed this classification in a
discrete binary fashion using $kd$ trees; see \citet{grn+05} and
\citet{rgr+08}.  However, we compute the continuous probabilities for
all of the objects that satisfy the initial binary selection criterion
and we report those values in the final catalog as they can sometimes
be useful in post assessment of the classification accuracy.

This process identified 1,317,677 objects as low-$z$ quasar
candidates, 804,342 as mid-$z$ quasar candidates, and 48,324 objects
as high-$z$ quasar candidates.  These amount to 2.6\%, 1.6\% and
0.1\% of the test set objects.  These percentages are larger than
expected from the priors; however, these include contaminants that we
have worked to remove using some cuts as described below and the
algorithm is not strongly sensitive to differences at this level.

We have reduced the amount of contamination from stars and galaxies by
restricting our analysis to objects classified as point sources in the
SDSS photometry and by requiring that all the candidates lie to the
right (redward) of {\em both} of the following two cuts:
\begin{equation}
([4.5]\le16.0 \&\& [3.6]-[4.5]< -0.1) || ([4.5]>16.0 \&\& [3.6]-[4.5]<([4.5]-15.2)/-8.0))
\end{equation}
\begin{equation}
(i\le16.5 \&\&  [3.6]-[4.5]<-0.1) || (i>16.5 \&\& [3.6]-[4.5]<(i-15.7)/-8.0).
\end{equation}
We further restrict our candidates to objects more than 15 degrees
from the Galactic plane and that have less than 1 magnitude of
extinction in the $u$-band, $A_u<1.0\;(A_i<0.4$).

After these cuts we are left with 
885,503 quasar candidates, including 748,839 low-$z$ candidates,
205,060 mid-$z$ candidates, and 13,060 high-$z$ candidates, where the
totals do not match due to objects being selected in more than one
redshift range.  These numbers can be contrasted with the 5546 quasar
candidates from our previous attempt at optical+MIR classification
\citep{rdl+09}.  Four of the mid-$z$ objects and five of the low-$z$
objects are duplicates that result from matching of multiple IR
sources to the same optical source; we have not resolved these
duplicates into a single object in the interest of completeness.

\begin{figure}[h!]
\plotone{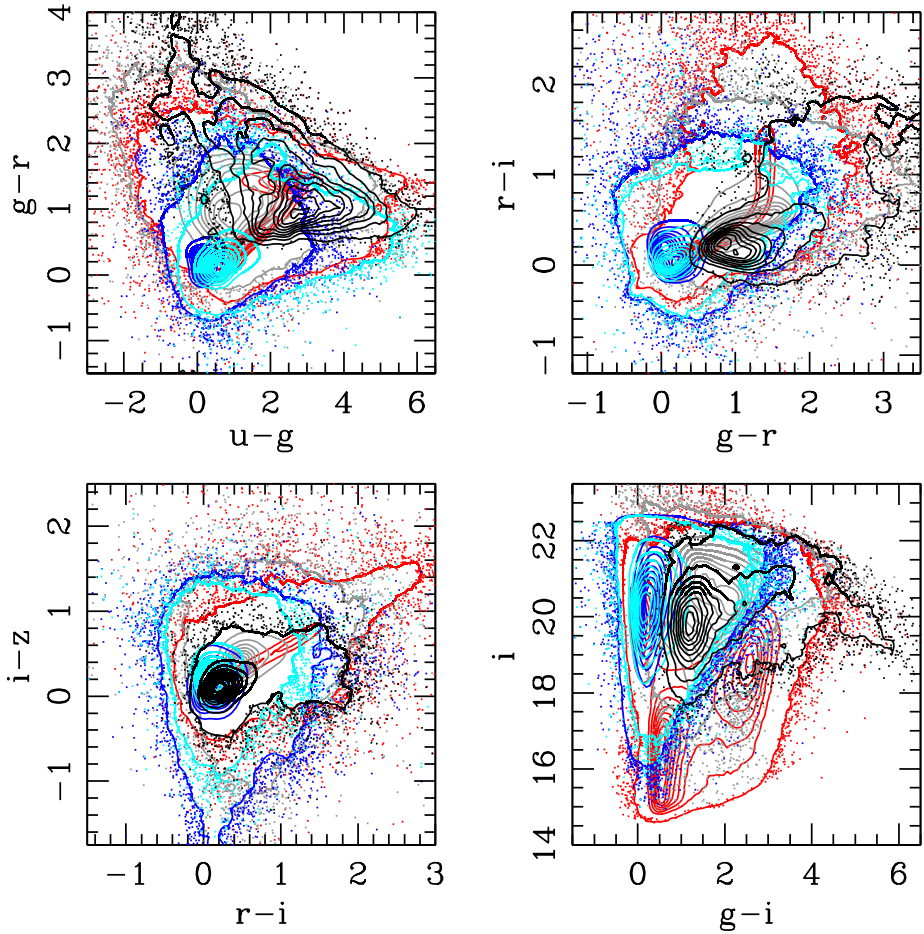}
\caption{Optical colors of test set objects selected as quasar candidates.  Contours/points and colors are as in Figure~\ref{fig:fig5}: high-$z$ quasars in black, mid-$z$ quasars in cyan, and low-$z$ quasars in blue.  
Training set ``stars'' are shown in red (for point sources) and gray (for extended sources).  
\label{fig:fig8}}
\end{figure}

\begin{figure}[h!]
\plotone{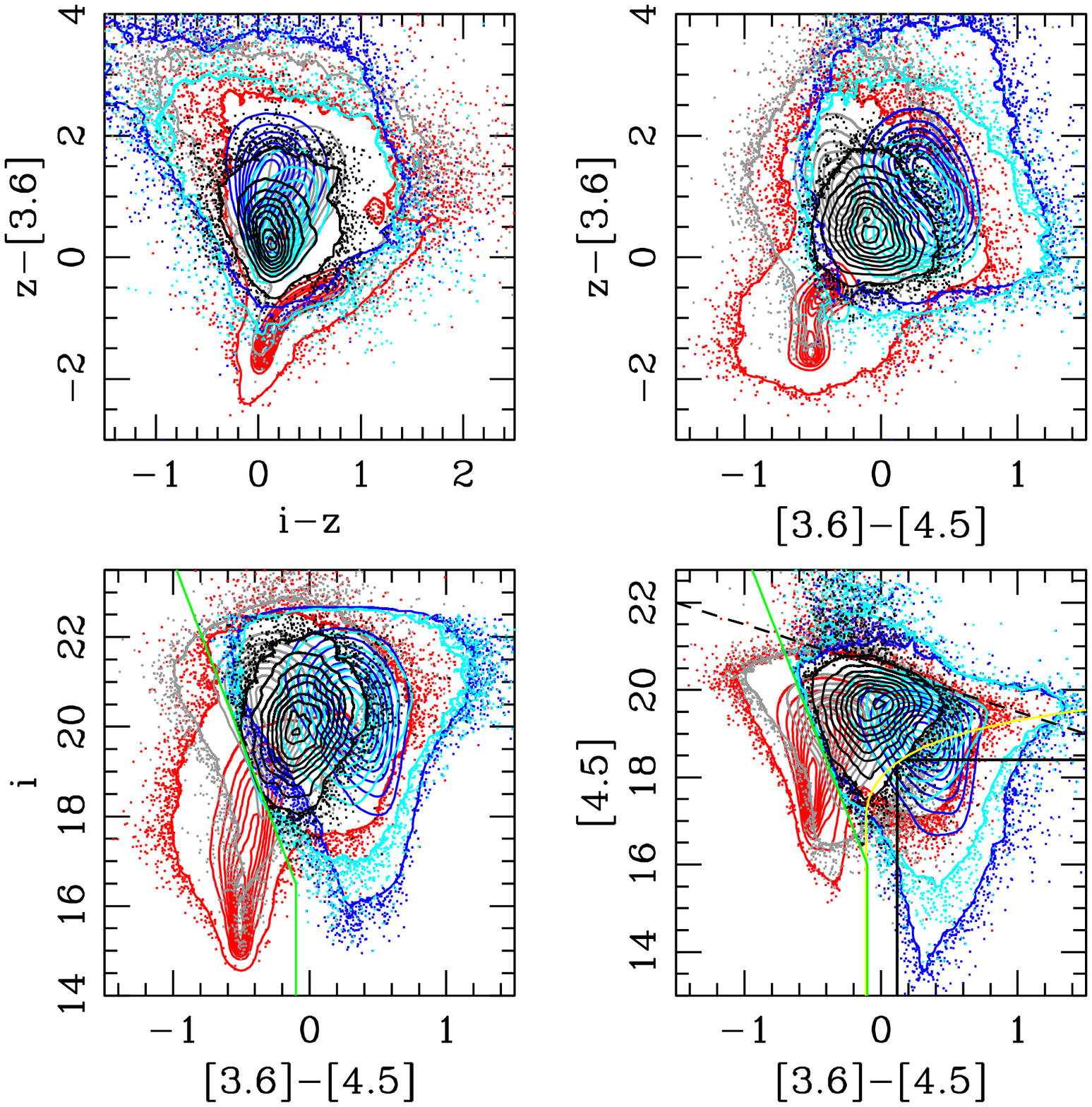}
\caption{Optical colors of test set objects selected as quasar candidates. Contours/points and colors are as in Figures~\ref{fig:fig6} and \ref{fig:fig8}.  The dashed black line shows the detection limit as a function of color for a theoretical object with [3.6]=20.5.  The solid black lines indicate the color and magnitude limits of the \citet{sab+12} selection in AB magnitude space, while the yellow curve gives the 75\% reliability selection from \citet{ask+13}.  The green lines in the bottom panels give our own cuts, as defined in Equations~3 and 4, that are intended to reduce stellar contamination. 
\label{fig:fig9}}
\end{figure}

Figures~\ref{fig:fig8} and \ref{fig:fig9} mimic
Figures~\ref{fig:fig5} and \ref{fig:fig6}, but
here we plot the quasar candidates rather than the known quasars.
Comparison of these distributions to the cuts used by \citet{sab+12}
(solid black lines in Fig.~\ref{fig:fig9}) and \citet{ask+13}
(solid yellow lines in Fig.~\ref{fig:fig9}) demonstrates the
improvement of our method to type-1 quasars (particularly those that
are faint with red optical colors) over using MIR color-magnitude cuts alone.  While
this vastly increases the number of high-$z$ quasar candidates, it
does come at the cost of excluding type 2 quasar candidates.

\section{The Catalog}
\label{sec:cat}

Our catalog is presented in Table~\ref{tab:cat}.  Of the 885,503
quasar candidates, 733,713 lack spectroscopic confirmation (and
305,623 are objects that we have not previously classified as
photometric quasar candidates).  We find that 150,453 objects are
already known to be quasars.  This was determined by matching the
candidates not only to the known quasars in the master quasar catalog
that defined our training set but also to the full SDSS-I/II/III
spectroscopic database.  Only 743 candidates ($<0.1$\%) have been
classified as stars.  A total of 589 objects are classified as
galaxies, however, 175 of those have $z>0.5$ and thus are likely to be
AGNs.  Indeed many of the objects classified as $z>0.5$ galaxies
appear in the hand-vetted SDSS quasar catalogs; this reflects the
sensitivity of our method to low-luminosity AGNs in compact galaxies.
The confirmed stars and galaxies do not occupy any unique parameter
space that would allow them to be easily distinguished as
contaminants.  Overall, the candidate list appears to be quite robust
and visual inspection of the optical SDSS images confirms this
impression.

\begin{deluxetable}{lll}
\tabletypesize{\scriptsize}
\tablewidth{0pt}
\tablecaption{Optical+MIR Photometric Quasar Catalog\label{tab:cat}}
\tablehead{
\colhead{Column} &
\colhead{Name} &
\colhead{Description} \\
}
\startdata
1 & RA & Right Ascension (J2000) \\
2 & DEC & Declination (J2000) \\
3 & CLASS & Spectral classifcation (QSO, GALAXY, STAR, CELG, ??, or U) \\
4 & ZSPEC & Spectroscopic redshift (if known) \\
5 & U\_MAG & SDSS $u$-band AB magnitude, corrected for Galactic extinction \\
6 & G\_MAG & SDSS $g$-band AB magnitude, corrected for Galactic extinction \\
7 & R\_MAG & SDSS $r$-band AB magnitude, corrected for Galactic extinction \\
8 & I\_MAG & SDSS $i$-band AB magnitude, corrected for Galactic extinction \\
9 & Z\_MAG & SDSS $z$-band AB magnitude, corrected for Galactic extinction \\
10 & CH1\_MAG & 3.6 micron AB magnitude, corrected for Galactic extinction \\
11 & CH2\_MAG & 4.5 micron AB magnitude, corrected for Galactic extinction \\
12 & U\_MAG\_ERR & Error on $u$-band magnitude \\
13 & G\_MAG\_ERR & Error on $g$-band magnitude \\
14 & R\_MAG\_ERR & Error on $r$-band magnitude \\
15 & I\_MAG\_ERR & Error on $i$-band magnitude \\
16 & Z\_MAG\_ERR & Error on $z$-band magnitude \\
17 & CH1\_MAG\_ERR & Error on 3.6 micron magnitude \\
18 & CH2\_MAG\_ERR & Error on 4.5 micron magnitude \\
19 & U\_FLUX & SDSS $u$-band flux density in nanomaggies \\
20 & G\_FLUX & SDSS $g$-band flux density in nanomaggies \\
21 & R\_FLUX & SDSS $r$-band flux density in nanomaggies \\
22 & I\_FLUX & SDSS $i$-band flux density in nanomaggies \\
23 & Z\_FLUX & SDSS $z$-band flux density in nanomaggies \\
24 & CH1\_FLUX & 3.6 micron flux density in microJy \\
25 & CH2\_FLUX & 4.5 micron flux density in microJy \\
26 & U\_FLUX\_ERR & Error in $u$-band flux density \\
27 & G\_FLUX\_ERR & Error in $g$-band flux density \\
28 & R\_FLUX\_ERR & Error in $r$-band flux density \\
29 & I\_FLUX\_ERR & Error in $i$-band flux density \\
30 & Z\_FLUX\_ERR & Error in $z$-band flux density \\
31 & CH1\_FLUX\_ERR & Error in 3.6 micron flux density \\
32 & CH2\_FLUX\_ERR & Error in 4.5 micron flux density \\
33 & YAPERMAG3 & $Y$-band Vega magnitude from UKIDSS or VHS \\
34 & JAPERMAG3 & $J$-band Vega magnitude from UKIDSS or VHS \\
35 & HAPERMAG3 & $H$-band Vega magnitude from UKIDSS or VHS \\
36 & KSAPERMAG3 & $K$-band Vega magnitude from UKIDSS or VHS \\
37 & YAPERMAG3ERR & Error in $Y$-band magnitude \\
38 & JAPERMAG3ERR & Error in $J$-band magnitude \\
39 & HAPERMAG3ERR & Error in $H$-band magnitude \\
40 & KSAPERMAG3ERR & Error in $K$-band magnitude \\
41 & FUV\_MAG & GALEX FUV magnitude (AB) \\
42 & FUV\_MAG\_ERR & GALEX NUV magnitude (AB) \\
43 & NUV\_MAG & Error in FUV magnitude \\
44 & NUV\_MAG\_ERR & Error in NUV magnitude \\
45 & GI\_SIGMA & Indicator of distance from mean g-i color at ZHOTBEST \\
46 & EXTINCTU & Extinction in SDSS $u$-band \\
47 & STAR\_DENS & Star Density from KDE algorithm \\
48 & QSO\_DENS & Quasar Density from KDE algorithm \\
49 & ZPHOTMIN & Minimum photometric redshift (ugriz) \\
50 & ZPHOTBEST & Best photometric redshift (ugriz) \\
51 & ZPHOTMAX & Maximum photometric redshift (ugriz) \\
52 & ZPHOTPROB & Probability of ZPHOTBEST being between min and max \\
53 & ZPHOTMINJHK & Minimum photometric redshift (ugrizJHK) \\
54 & ZPHOTBESTJHK & Best photometric redshift (ugrizJHK) \\
55 & ZPHOTMAXJHK & Maximum photometric redshift (ugrizJHK) \\
56 & ZPHOTPROBJHK & Probability of ZPHOTBESTJHK being between min and max \\
57 & LEGACY & Indicates if object is in the SDSS Legacy footprint \\
58 & SDSS\_UNIFORM & Indicates if object was selected according to Richards et al. (2002) \\
59 & PRIMTARGET & SDSS primary target selection flag; see Richards et al. (2002) \\
60 & PM & Proper motion in milliarcseconds per year \\
61 & DUPBIT & Bitwise flag indicating low-$z$ ($2^0$), mid-$z$ ($2^1$), and high-$z$ ($2^2$) sources 
\enddata
\end{deluxetable}

The columns in the catalog are as follows.  1) RA (degrees; J2000), 2)
Declination (degrees; J2000), 3) the classification of the object from
matching to known objects (QSO, STAR, GALAXY, CELG, and
``??'')\footnote{See Section~\ref{sec:cosmos} for an explanation of
  the ``CELG'' and ``??''  classifications.} based on existing
spectroscopy, or ``U'' for unknown if we know of no spectroscopy for
the source, and 4) the known redshift.  Columns 5-11 give the $ugriz$
optical AB (asinh) magnitudes (corrected for Galactic extinction)
along with the $[3.6]$ and $[4.5]$ mid-IR AB magnitudes (also
corrected for Galactic extinction).  Columns 12-18 give the errors on
these magnitudes.  Columns 19-32 give the SDSS-III, {\em WISE}, and
{\em Spitzer} flux densities and errors where the optical values are
measured in nanomaggies (as reported by the SDSS data sweeps file that
we use) and the mid-IR values have been converted to $\mu$Jy; no
extinction correction is applied to these values.  Columns 33-40 give
the $YJHK$ magnitudes and errors from the UKIDSS or VHS surveys (where
available).  Columns 41-44 give the far-UV and near-UV magnitudes
  and errors from {\em GALEX} (where available); no Galactic
  extinction corrections have been applied.  Column 45 indicates
whether the $g-i$ color is within 1$\sigma$ (0.68), 2$\sigma$ (0.95),
or 3$\sigma$ (0.99) of the mean color for quasars at the predicted
redshift.  Outliers are an indication of either bad photometric
redshifts or non-quasar contaminants.  Column 46 is the $u$-band
extinction from \citet{sf11}; extinctions in other wavebands can be
derived from this value.  Columns 47 and 48 are the star and quasar
probabilities as determined by the kernel density estimation used in
our primary selection criterion.  Columns 49-52 use the optical and
MIR photometry to tabulate the minimum, best, and maximum photometric
redshift along with the probability of being between the minimum and
maximum values as described in more detail in
Section~\ref{sec:photoz}.  Columns 53-56 are the same photometric
redshift values but now adding $JHK$ photometry to the optical and
MIR.  Column 57 indicates whether the object is within the ``legacy''
SDSS footprint, which is useful for statistical analysis.  Column 58
indicates whether the objects was in the uniform targeting area for
the quasar target selection algorithm described in \citet{rfn+02}.
In Column 59 we give the flag (if set) from SDSS-DR7 quasar
  targeting, where \citet{rfn+02} and \citet{srh+10} provide details
  on the values of these flags---which can be used as a secondary
  indicator of quasar likelihood.  Column 60 gives the proper motion
  ({\tt PM}) in mas per year in a similar manner as discussed in
  \citet{rmg+09}, based on \citet{mml+04} and can also be used as a
  secondary indicator of quasar likelihood.  Finally, column 61 is a
bit-wise flag that indicates whether the object was selected as a
low-$z$ ({\tt DUPBIT} \& $2^0$), mid-$z$ ({\tt DUPBIT} \& $2^1$), or
high-$z$ ({\tt DUPBIT} \& $2^2$) source (or a combination thereof).

\subsection{Photometric Redshifts}
\label{sec:photoz}

We have used the photometric-redshift algorithm described by
\citet{rws+01} and \citet{wrs+04}, extending it to include the mid-IR
photometry from {\em Spitzer} and {\em WISE}, and, in some cases, the
near-IR photometry from VHS and UKIDSS.  In short, this algorithm
seeks to minimize the distance between the colors of an unknown source
and the mean colors as shown in Figure~\ref{fig:fig4}.  For luminous
quasars this method is superior to template fitting
\citep[e.g.,][Fig.~10]{akb+10} as it primarily picks up on the
high-equivalent-width emission line features rather than spectral
breaks (although at high-$z$ the Ly$\alpha$ break leads to improved
photometric redshifts even with our method).  Careful selection of
templates can lead to improved results as shown by
\citet{shi+09}---particularly for host-dominated AGNs.

\begin{figure}[h!]
\plottwo{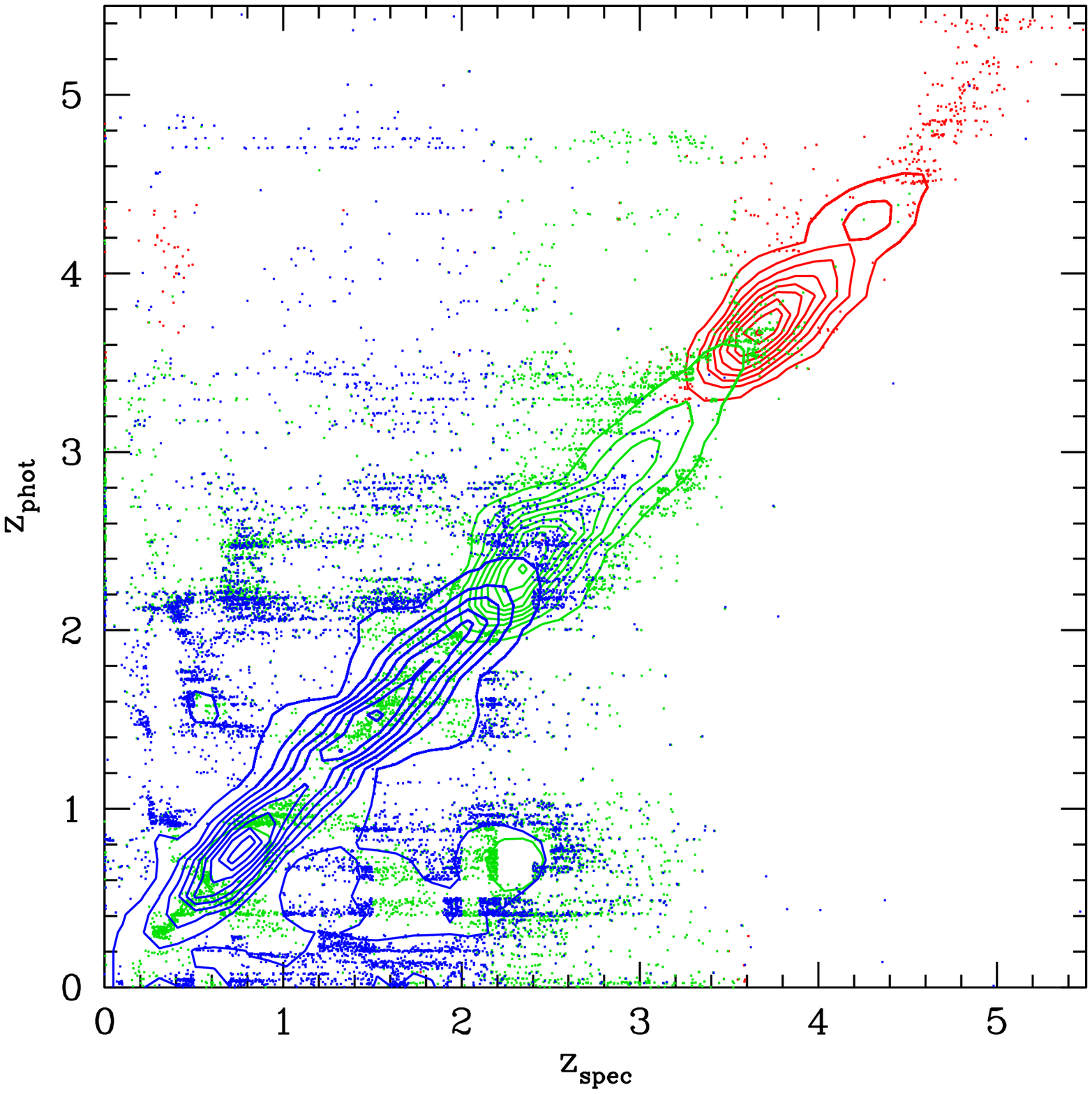}{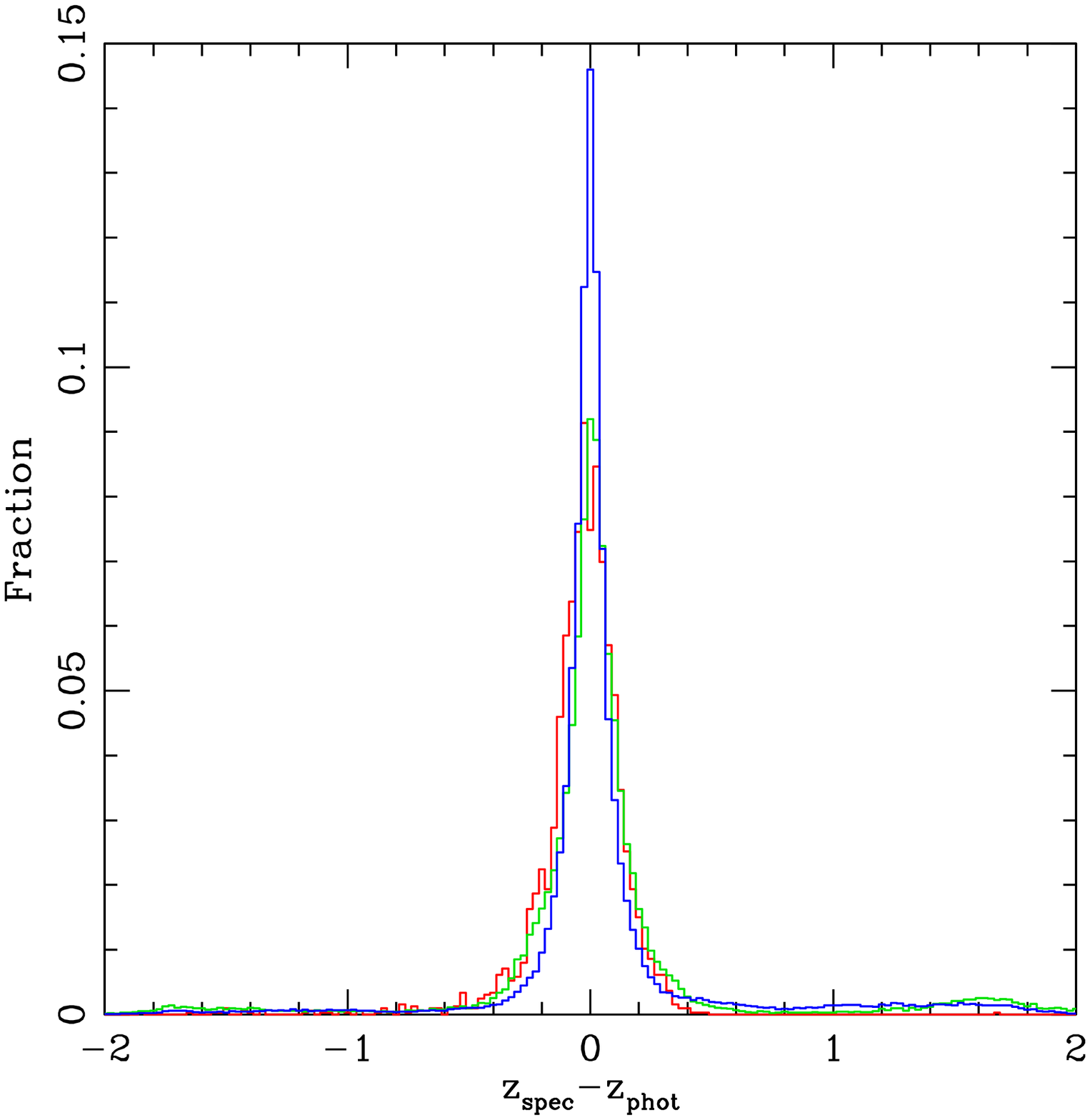}
\caption{(Left:) Photometric vs.\ spectroscopic redshift for all 3 samples; blue: low-$z$, green: mid-$z$, red: high-$z$.  As this presentation highlights the catastrophic outliers at the expense of the well-determined photometric redshifts we also present a histogram of the differences between the spectroscopic and photometric redshifts in the {\em right} panel.   This shows that most objects have well-estimated redshifts.
\label{fig:fig10}}
\end{figure}

Figure~\ref{fig:fig10} shows the photometric vs.\ spectroscopic
redshifts for all three samples.  Note that there is some overlap
between the samples (as designed to ensure that objects with redshifts
near the edges of the training set redshift windows are not lost).
The left panel reveals where there are photometric redshift
degeneracies in the sample; however, the right panel shows that
the vast majority have well-estimated photometric redshifts and that
catastrophic outliers are a minority.  We find that 90.9\%, 82.7\% and
85.7\% of known quasars have photometric redshifts within $\delta
z=\pm0.3$, for high-$z$, mid-$z$, and low-$z$ candidates,
respectively.  Candidates can be restricted to more robust photometric
redshifts by making a cut on {\tt ZPHOTPROB} which gives the
probability that the true redshift is between the minimum and maximum
reported values.

It is not our goal herein to rigorously investigate the nature of the
degeneracies in Figure~\ref{fig:fig10}.  However, as one example, we
consider the degeneracy between $z\sim0.75$ and $z\sim2.25$.  Here the
Lyman-$\alpha$ forest is not yet strong enough in $u$ to overcome
similarities between the general optical/UV and MIR spectral slopes,
\ion{Mg}{2} vs.\ \ion{C}{4} in $g$, H$\beta$ vs.\ \ion{Mg}{2} in $z$,
and Pa$\alpha$ vs.\ Pa$\gamma$ in $[3.6]$.  $JHK$ data can break that
degeneracy as $J-K$ spans the 1\,$\mu$m transition between the optical
and IR at low redshift while it samples the optical slope at high
redshift.  We specifically find that adding $JHK$ data improves the
overall photo-$z$ accuracy to 93\% (virtually eliminating catastrophic
errors).  However, near-IR data of sufficient depth are only available
over a fraction of the area surveyed; {\em Euclid} data
  \citep{Euclid} will be very welcome in this regard.

\begin{figure}[h!]
\epsscale{0.6}
\plotone{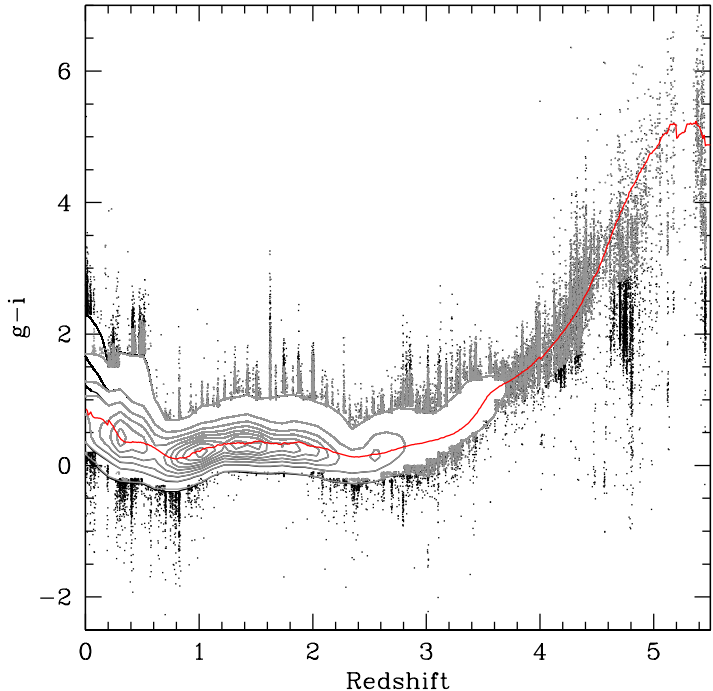}
\caption{$g-i$ color vs.\ {\em photometric redshift} for our quasar candidates.  Black (linear) contours and dots are all the candidates; gray contours and dots represent objects that have colors that are within the 99\% confidence limits of the mean quasar color-redshift relation ({\em red line}).  Outliers may be contaminants or have erroneous photometric redshifts.
\label{fig:fig11}}
\end{figure}

Another way we can determine the photometric redshift accuracy is to
look at the color-redshift relation using the photometric redshifts of
our objects.  Figure~\ref{fig:fig11} shows the distribution of $g-i$
color versus {\em photometric redshift} for our candidates.  Photometric
redshift degeneracies can produce semi-discrete features where one
redshift is preferentially selected.  Objects where the $g-i$ color is
within the 99\% confidence limit at the best-fit photometric redshift
({\tt GI\_SIGMA}) are highlighted in gray.  These objects are likely to be
the most robust candidates and are expected to have the most accurate
photometric redshifts.  Objects outside of this 99\% confidence
interval are likely to be contaminants, have erroneous photometric
redshifts or have interesting spectral features (highly dust reddened,
broad absorption lines, etc.).  For example, the objects with colors
bluer than the mean $g-i$ color at photometric redshifts of $z\sim4.8$
and $z\sim5.5$ are unlikely to be at those redshifts.  However, they
may well be quasars at $z\sim4$--$4.5$.  Alternatively, if they are
indeed quasars, they could be at much lower redshift but have dust
reddening or absorption troughs that make them appear like higher
redshift quasars.


\section{Analysis}
\label{sec:analysis}

\subsection{Comparison of Selection Methods}

An advantage of our selection method is that it can take full
advantage of data from {\em Spitzer} during the post-cryogen
exploration phase of the mission.  In such cases, we only have
3.6$\mu$m and 4.5$\mu$m measurements.  This keeps us from being able
to perform ``wedge'' selection that has proven so successful
\citep{lss+04,seg+04,ska+07} because {\em WISE} is not deep enough in
$W_3$ and $W_4$ relative to our optical data.
However, our method allows us to probe to much fainter IR limits using only 2
bands since we also have matched optical photometry.  This process
enables us to improve upon MIR-only selection \citep{sab+12,ask+13}
(at least within the SDSS footprint) by helping to remove the MIR bias
against $3.5<z<5.0$ quasars.  In a similar vein, our method is
potentially more complete and more efficient at faint magnitudes than
variability selection \citep[e.g.,][]{bb11}, where the optical
photometry in any single epoch of imaging is noisy.  While the
power-law method used by \citet{dkb+12} results in more reliable MIR
classification, quasars are not necessarily power-laws in the MIR (and
are not always monotonic), so that method is more incomplete
than that presented herein with regard to those objects that do not
fit a power-law template (to within the errors).

We note that the \citet{bhh+11} algorithm should perform similarly
well as ours if it were rebuilt to include the stellar locus in the
MIR (as opposed to applying color cuts before/after running the
optical selection algorithm).  One utility of the \citet{bhh+11}
method is that meaningful probabilities can be easily and rapidly
built on a per-object basis. This allows for the alternative approach
of constructing a fully probabilistic or {\em extremely complete}
catalog, which is a less appropriate catalog to use for direct
statistical analyses but which can be used to, e.g., match
low-probability objects in the optical+MIR to AGN selected at other
wavelengths (see DiPompeo et al., 2015, in preparation, for just such
a catalog).  Alternatively, the catalog we have built is deliberately
{\em efficient} (or ``pure'') and therefore more appropriate for
statistical analyses given good characterization of the incompleteness.

An obvious question is what our method has gained over making simple
color cuts.  We illustrate this with two examples of MIR-only cuts and
a cut involving both optical and MIR data.
\citet[][Figure~10]{rmg+09} illustrates the trade-off between
completeness and contamination for a simple $[3.6]-[4.5]$ color-cut.
The standard $W_1-W_2>0.8$ cut, which equates to $[3.6]-[4.5]>0.119$
as discussed above would recover 80\% of the quasar candidates
compiled herein, with most of the losses being high-redshift
candidates.  The total number of test set objects passing such a cut
is 1.85M.  If all of our candidates were quasars and all of the
remaining objects within those 1.85M were contaminants, then the
contamination of such a cut would be 60\%.  Restricting just to point
sources leaves only 1M targets, but that still would represent a
contamination of 30\%.  Thus such a cut would neither be optimally
complete or efficient.  If we wanted a more complete quasar set, a
better cut would be $[3.6]-[4.5]>-0.1$, which achieves 95\%
completeness to our quasar candidates.  However, it obviously comes
with significantly greater contamination: 86\% overall and 55\% for
point sources.

Better yet would be to combine the optical and MIR color information
as we have in our KDE selection.  A number of combinations are
possible, but a simple cut of $i-[4.5]>(g-i) - 1$ recovers 99\% of our
candidates.  With that comes nearly 95\% contamination as more than
18M other objects are also selected by this cut.  Most of that
contamination is from normal galaxies as restricting to point sources
reduces the contamination to 50\%.  A more restrictive cut to reduce
the contamination is possible, but not without a commensurate
reduction in completeness.  

\subsection{Creating Robust Subsamples}
\label{sec:robust}

In order to further compare our candidates and selection algorithm to
others, it is helpful to first identify the most robust subsamples
possible.  To that end we consider the effects of star-galaxy
separation, previous SDSS targeting flags, proper motion, and the
presence of {\em GALEX} detections.

Particularly at high-$z$ the robustness of our candidates depends on
SDSS star-galaxy separation (as we might expect high-$z$ quasars to be
point sources).  The morphological classification is thought to be
95\% correct at $r\sim21$ \citep{Annis11}, where this has been
explored in more detail by \citet{sjd+02}.  Figure 1 of \citet{sjd+02}
shows that, as S/N degrades, galaxies are more likely to have small
concentration indices and thus be classified as stars.  ``Point''
sources fainter than 22nd mag have a significant probability of being
galaxies; in poor seeing it is closer to 21st mag.  As such, we do not
consider any $i>22$ sources to be robust high-$z$ candidates (in the
absence of other confirming information) and sources with $21<i<22$
deserve some caution.

In the case of relatively bright sources, the \citet{rfn+02} SDSS
quasar target selection flags can be used to identify candidates that
are particularly likely (or unlikely).  As such, we have included
those target flags (in the field {\tt PRIMTARGET}) for sources where
the SDSS-DR7 flag value was non-zero.  Objects flagged as QSO\_FAINT
({\tt PRIMTARGET} \& 0x02000000) are sources that otherwise met the
SDSS-DR7 selection criteria, but were just below the flux limit for
spectroscopic follow-up.  On the other hand, objects flagged as
QSO\_REJECT ({\tt PRIMTARGET} \& 0x20000000) are in regions of color
space known for high contamination.  Based on the known quasars and
the color cuts that defined this flag, objects with this flag set that
do not have $z_{\rm phot}\sim2.4$ are likely to be less robust
candidates.

In \citet{rmg+09} we were able to remove some contaminants by
identifying objects with high proper motions \citep{mml+04} and we
have included the proper motion for those objects with quality proper
motion measurements (having small errors and at least 6 epochs of
data; see the discussion in \citealt{rmg+09}).  Using same cuts as
\citet{rmg+09} removes 160 known quasars which is just 0.25\% of the
quasars with quality proper motion measurements, yet it cuts 59 of the
280 (21.1\%) of the known stars.  These criteria further cut 478
unknown objects (0.73\%) as compared to the 163 expected if all of
those objects were quasars.  Overall, we find that many fewer objects
have large proper motion than in \citet{rmg+09}, which we attribute to
the current catalog being less contaminated by stars.  

We have not used UV data from {\em GALEX} in our selection or
photometric redshift analysis, but we have further matched our catalog
to {\em GALEX} data in order to identify contaminants and redshift
errors.  Specifically we matched our candidate quasars to both the MIS
and AIS {\em GALEX} catalogs as compiled by \citet{bsz+11}, excluding
sources with an near-UV (NUV) artifact flag.  We then tabulate the NUV
and FUV (far-UV) magnitudes (AB) in addition to their errors.  This
matching can be used to weed out low-$z$ interlopers from among the
high-$z$ candidates.  Specifically, real high-$z$ quasars are
relatively unlikely to be {\em GALEX} sources (particularly fainter
sources).  Alternatively, lower-redshift sources that we have
misclassified as high-$z$ quasar candidates are much more likely to be
detected by {\em GALEX} in the UV.  We find that 101 of 9283 (1.1\%)
known quasars in our sample with $z_{\rm spec}>3$ are detected by {\em
  GALEX}, as compared to 313 of the 9547 objects (3.3\%) with $z_{\rm
  phot}>3$, but that have low probability ($<0.8$) of being at $z>3$.
Thus a {\em GALEX} detection for a high-$z$ candidate suggests that
the candidate may not be robust.

The end result of these investigations is the addition of a number of
parameters to our catalog that can be used to identify the most robust
candidates.  For our purposes, we will formally define ``robust''
candidates as those having ${\tt ZPHOTPROB}>0.8$ and $abs({\tt
  GI\_SIGMA})<=0.95$.  There are 517586 candidates satisfying these
criteria.  Of those only 717 (0.14\%) are known non-quasars, whereas
114120 are known quasars.

For high-$z$ candidates ($3.5<z<5$) we further restrict the most
robust sources to non-detections in {\em GALEX} and $i<22$.  There are
10955 such sources, of which 7874 are unknown; 6779 of these have not
been previously identified by us as photometric quasar candidates.
Only 79 are non-quasar contaminants, while 2890 of the 3002 known
quasars (96\%) indeed have $z>3$.

\subsection{COSMOS and Bootes}
\label{sec:cosmos}

One way to judge the utility of this catalog is to compare it to areas
for which there is particularly dense spectroscopy.  One such example
is the COSMOS field \citep{SCOSMOS}.  In addition to the COSMOS
spectroscopy discussed in Section~\ref{sec:master}, we also compared
to \citet{pic+06}, which further identifies objects in the COSMOS
field.  We recover 75 of the 95 quasars cataloged by them.  Thirteen
of these 75 were not identified as quasars in the master catalog and
we have updated their classifications in our catalog.  Only 3 of our
objects match to galaxies from \citet{pic+06} while no objects matched
to stars.  

This comparison suggests that our catalog is relatively complete to
known COSMOS quasars and has relatively little contamination.  Yet our
catalog has nearly as many new quasar candidates within the COSMOS
field as have been confirmed by spectroscopy.  Within the area bounded
by the COSMOS {\em Spitzer} data, we find 547 quasar candidates in
total.  Of these 266 are known quasars, 3 are known galaxies, 1 is a
known star, 32 are known compact emission line galaxies (CELG), 5 have
spectra that are difficult to classify (given as ``??'' in the
catalog), and 240 are unknown.  CELG is a designation that we have
chosen for those objects that are classified as narrow line in the
COSMOS spectroscopy but generally show signs of being star forming
galaxies rather than being AGN powered.  They are all fainter than
$i=21$ and likely come into the sample due to a breakdown of SDSS
star-galaxy separation as noted above.  Of the unknown objects, only
95 are robust candidates as described above (20 with $z_{\rm
  phot}>3.5$ and $i<22$).  The lower-quality candidates have $i\sim22$
and are at the limit of our selection method.  Of the known quasars,
only 5 have $z>3.5$ and 47 have $2.2 \le z \le3.5$.

We can further compare our candidates to X-ray sources in the COSMOS
field.  The November 2011 update of the 53-field {\em XMM-Newton}
data table analyzed in \citet{bcc+10} contains 2000 X-ray sources.
There are 264 matches (to within 1$\arcsec$) to our catalog, 28 of
which are unknown (16 robust).  However, there are 176 additional
unknown candidates (64 robust) from our catalog without X-ray matches
that we deem within the X-ray footprint by virtue of there being an
X-ray source within 240$\arcsec$ (i.e., they are quasar candidates but
were not detected in the X-ray).  Of the robust candidates, 17 are
$z>3.5$ candidates with $i<22$.
Comparing the candidates matched to X-ray sources and those not
matched we find that the average $i$-band magnitude of the matches is
20.62, while for the non-matches it is 21.96.  In terms of photometric
redshift, the X-ray matches have a mean value of 1.28 as compared to 2.88
for the non-matches.

{\em Chandra} data in COSMOS cover a slightly smaller region.  Using
the Chandra Source Catalog\footnote{http://cxc.harvard.edu/csc/} we
find 934 X-ray sources of which 125 match to our candidates with 3 of
those being objects without existing spectroscopy.  However, there are
another 125 of our quasars candidates within this X-ray footprint.  20
of those are robust unknown sources with 7 that are $z>3.5$ candidates
with $i<22$ (all of which are included in the {\em XMM} matching
above).  The average magnitude for these X-ray matches is $i=20.75$
and for the non-matches is $i=21.54$.  The mean photometric redshift
for matches is $z=1.24$ and for non-matches is $2.69$.

In principle, we could use morphology to further test the likelihood
of the quasar classification of our the candidates.  However the SDSS
star-galaxy separation becomes unreliable at a brighter limit than our
candidates.  Although deep {\em HST} data are available in the COSMOS
area \citep{saa+07}, it is not definitive.  While the known bright
quasars do tend to have point-like morphologies, the faint quasars
(even at high-z) can be extended (host dominated) at the depth of the
{\em HST} data.
That said, any follow-up spectroscopy of
COSMOS candidates should clearly consider the {\em HST} data for
prioritization as 6 of the 12 new high-$z$ candidates have stellar
morphologies from {\em HST} (with the 5 non-matches to the {\em HST}
data all being near the edges of the COSMOS field).

If even a fraction of our mid- and high-$z$ quasars candidates
in the COSMOS area are real quasars, it would significantly increase
the number of such objects.  Compared to only 5 known $z>3.5$ quasars
among our candidates, we saw above that there are 17 robust $z>3.5$
candidates just within the X-ray footprint of the field, 6 of which
have stellar {\em HST} morphologies---suggesting that the existing
density of relatively bright high-$z$ quasars in the COSMOS field is
at least $\sim$50\% incomplete.

The photometric redshifts for COSMOS sources presented by
\citet{sih+11} should be superior to ours and can be used to
cross-check our results.  However, only 36 of our candidates match: 12
mid-$z$ and just 1 high-$z$, likely because of the restriction to
X-ray sources in \citet{sih+11}.  Of these, 17 have photometric
redshifts that agree with ours to within $\pm0.3$ (9 to $\pm0.1$),
including the high-$z$ candidate (COSMOS ID: 1980473) with a
photometric redshifts of 3.295 vs.\ 3.329.

\citet{bcg+09} report 40 $z>3$ quasars in the COSMOS field (22
spectroscopic, 18 photometric).  We recover only 7 of those (all of
which already appear in the master quasar catalog); however, this is
not surprising as, of the 33 missing, 32 have $i>20.5$ (the peak of
our distribution) and 27 have $i>22$, thus the \citet{bcg+09} objects
are much fainter than those cataloged herein.

As with the COSMOS field, the Bo\"{o}tes field has also been subject
to considerable spectroscopic exploration, primarily from the AGES
program \citep{kec+12}.  Within a rectangular area defined by the
minimum and maximum RA and Dec of the deep {\em Spitzer} data taken as
part of the Spitzer Deep, Wide-Field Survey (SDWFS; \citealt{SDWFS}),
we find 1861 quasar candidates.  Among these are 1085 confirmed
quasars, 2 stars, and 3 galaxies, leaving 771 unknown objects.
However, the {\em Spitzer} data do not fully cover this space: there
are 1738 candidates (of which 681 have no spectroscopic data) that are
included within the approximate boundaries of the MIR data.  Some of
those objects fall outside of the boundaries of the AGES spectroscopic
program \citep[][Figure~2]{kec+12}, but nevertheless have the deep MIR
data needed to perform robust MIR selection.

Matching back to the AGES spectroscopy (to recover non-quasars not
included in the training set), we find an additional 3
spectroscopically-confirmed stars and 36 spectra that resulted in
unknown redshift/classification.  A search of the NASA Extragalactic
Database for additional spectroscopic data revealed only one new
object: FBQS~J142607.7+340426 that was not included in our master quasar catalog.

Thus, as with the COSMOS field, the Bo\"{o}tes field also contains
many new quasar candidates, despite considerable efforts to confirm
likely AGN.  Of the 771 unknown candidates, we find that 294 are
robust, with 46 being robust $z>3.5$ candidates with $i<22$.

As a result of this analysis of quasar candidates in the COSMOS and
Bo\"{o}tes fields, we conclude that there is a potential for
significantly increasing the number of relatively-bright high-$z$ (and
mid-$z$) quasars in that area of sky---despite considerable existing
spectroscopic coverage of the field.  The density of objects in these
(and other {\em Spitzer} deep fields) is particularly useful for
absorption studies, making additional confirming spectroscopy
worthwhile.



\subsection{Demographics}


One of our goals was to fill in the gaps at redshifts where
optical-only quasar selection has traditionally been incomplete.  The
SDSS selection algorithm \citep{rfn+02} targets both low-redshift and
high-redshift quasars to $i<19.1$.  To that limit the SDSS quasar
sample is expected to be quite complete at $z<2.2$, with known
incompleteness at $z\sim2.7$ and $z\sim3.5$
\citep{vsr+05,rsf+06,wp11}.  Similarly the BOSS selection algorithm
\citep{rms+12} is limited to $\sim2.2<z<\sim3.5$ and has known
incompleteness at $z\sim2.9$ \citep{rmw+13}.  We would thus expect to
find that our method would have little new to offer in terms of new
bright quasars ($i<19.1$) at $z<2.2$, but may significantly improve
quasar selection around $z\sim2.7$ and $z\sim3.5$.  We might expect
somewhat more new quasar candidates between $19.1<i<20.2$ as SDSS did
not target quasars at $z<3$ fainter than $i=19.1$ (reserving the
fainter targets for $z>3$ candidates---targeted to $i<20.2$) and BOSS
did not explicitly target $z<2.2$ quasars.

\begin{figure}[h!]
\epsscale{0.8}
\plotone{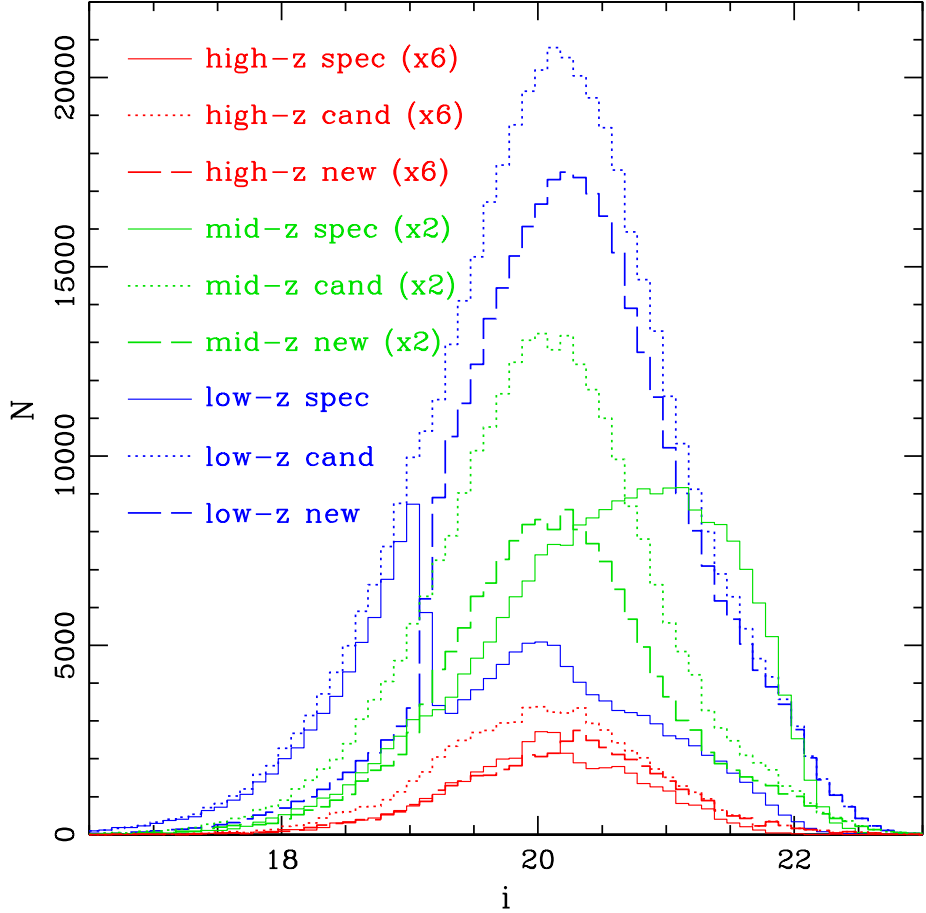}
\caption{Number counts of known quasars and robust quasar candidates as a function of magnitude and redshift range.  Blue lines show the number of known quasars with $z<2.2$ (``spec''; thin), the number of low-$z$ selected candidates (``cand''; dotted) and the number of low-$z$ selected candidates that lack spectroscopic confirmation (``new''; dashed).  Similarly green and red lines give the number of $2.2<z<3.5$ (or mid-$z$ selected) and $3.5<z<5.5$ (or high-$z$ selected) quasars and quasar candidates.  The green curves are scaled up by a factor of 2 and the red curves are scaled up by a factor of 6 in order to made the figure more legible.
\label{fig:fig12}}
\end{figure}

In this light, we have matched our candidate list to the full master
catalog (to determine which of these objects are new candidates), to
the training sets (to determine the completeness with respect to the
quasar training set), and to the full SDSS-III spectroscopic database
(to identify known non-quasars).  Figure~\ref{fig:fig12}
compares the number of known spectroscopic quasars, our robust quasar
candidates, and those robust candidates without existing spectroscopy.
Comparing the low-$z$ quasars/candidates (blue lines) we find that
there are some new quasars at $i<19.1$ (the SDSS spectroscopic limit
for $z \lesssim 3$), which may reflect our sensitivity to
low-luminosity AGNs in compact galaxies.  There are also hundreds of
thousands of new low-$z$ objects at fainter magnitudes.

For mid-$z$ selected quasars ($2.2<z<3.5$),
Figure~\ref{fig:fig12} shows that our catalog provides relatively
little in terms of new sources at $i<19.1$ and $i>21$\footnote{Note
  that the thin solid lines in Figure~\ref{fig:fig12} show the
  total number of known spectroscopic quasars, not the number of such
  objects that also have mid-IR photometry.  Thus even if the
  candidate object number counts are below the spectroscopic counts,
  that does not necessarily indicate that we are incomplete to known
  quasars with MIR data.}.  However, at intermediate magnitudes, the
number of new candidate mid-$z$ quasars is quite substantial.  In some
sense this is surprising as the SDSS-III BOSS project was specifically
designed to find quasars in this magnitude and redshift range.  At the
same, it is known that BOSS is only $\sim$60\% complete
\citep{rmw+13}, so it is quite possible that we are simply turning up
the remaining objects missed by BOSS.  For high-$z$ selected quasars
($z>3.5$), we again find relatively few new objects brighter than the
SDSS spectroscopic limit (here $i<20.2$), but there is a significant
population of new candidates at fainter magnitudes---consistent with
the difficulty that standard wedge-based IR selection of AGNs
\citep{lss+04,seg+04,ska+07} have to recover objects at these
redshifts.

\begin{figure}[h!]
\epsscale{0.8}
\plotone{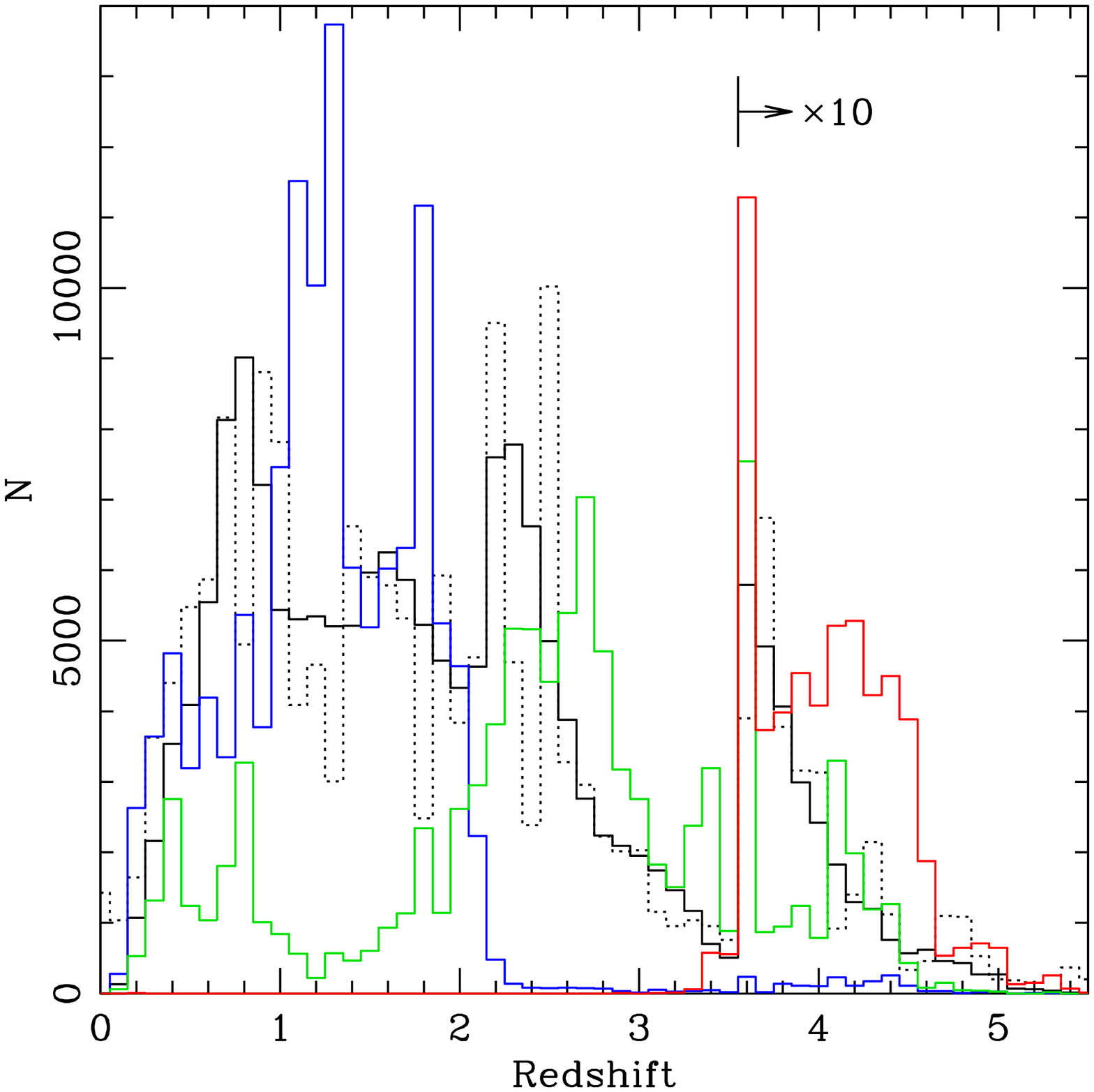}
\caption{Number of quasars as a function of redshift.  The solid black line gives the {\em spectroscopic} redshift distribution of the quasars in our training set, while the dotted black line gives the {\em photometric} redshift distribution for those same sources.  The ratio of these two is used to perform a first-order correction of the photometric redshift distribution of our candidates.  Corrected photometric redshift distributions for the robust new candidates (spectroscopically-confirmed sources removed) are shown in blue for low-$z$ candidates (scaled down by 3$\times$ to fit on the axis), green for mid-$z$, and red for high-$z$.  All histograms are scaled up by 10$\times$ for $z>3.6$ to better show the high-redshift distribution.
\label{fig:fig13}}
\end{figure}

The expected redshift distribution of the robust new candidates is shown in
Figure~\ref{fig:fig13}.  We have computed the ratio of the photometric
and spectroscopic redshift distributions for the spectroscopically
confirmed quasars in our training set.  This enables a rough
correction of the photometric redshift distributions of our candidates
to an expected spectroscopic redshift distribution (shown in blue,
green, and red for low-$z$, mid-$z$, and high-$z$ candidates,
respectively).  As noted above, the low-$z$ candidates are largely
faint sources; they generally have photometric redshifts consistent
with their low-$z$ selection.  The mid-$z$ candidates have a large
range of photometric redshifts, which suggests photo-$z$ degeneracy
and/or contamination.  There are a large number of mid-$z$ candidates
with photometric redshifts of $z\sim2.7$ and $z\sim3.5$, which is
encouraging as these are redshift regions where we know that
optical-only selection is incomplete \citep{rfn+02,rsf+06}.  The
high-$z$ candidates all have redshift estimates consistent with their
selection, with a large number of new objects spanning $3.6<z<4.6$.

We find that most of the new candidates are at fainter magnitudes
and/or come at redshifts where it is difficult to do optical-only,
variability-only, or infrared-only selection of quasars.  For example,
many new candidates are at high-redshift which tend to be biased
against by traditional mid-IR selection methods as noted above and
also by variability selection methods.  Overall, there are 7874 robust
high-$z$ quasar candidates.  If all turned out to be quasars, this
would more than double the number of such quasars in the SDSS
footprint.  Many of these candidates are very faint, but the
distribution peaks at $i\sim20.5$, likely reflecting the cutoff of
$i=20.2$ for high-$z$ quasars selection in SDSS-I/II.
In the mid-$z$ range there are 81,321 robust quasar candidates. 
At low-$z$ there are 424,448 robust quasar candidates.  Most of these are
quite faint, and despite the catalog's limitation to point sources, those with
$z_{\rm phot}<1$ are likely AGNs rather than luminous quasars.

Many of these candidates are identified in our previous photometric
quasars catalogs.  However, a total of 87,242, 34,059, and 6779
low-$z$, mid-$z$, and high-$z$ candidates respectively do not already
appear in \citet{rmg+09} or \citet{bhh+11}.


\subsection{Number Counts/Luminosity Function}

A particularly useful test for a sample of photometric quasars is a
comparison of their number counts to those of known quasars.  Problems
with efficiency/contamination will show up as an excess of quasars
(particularly at bright magnitudes), while problems with completeness
will show up as a dearth of quasars.  

\begin{figure}[h!]
\plotone{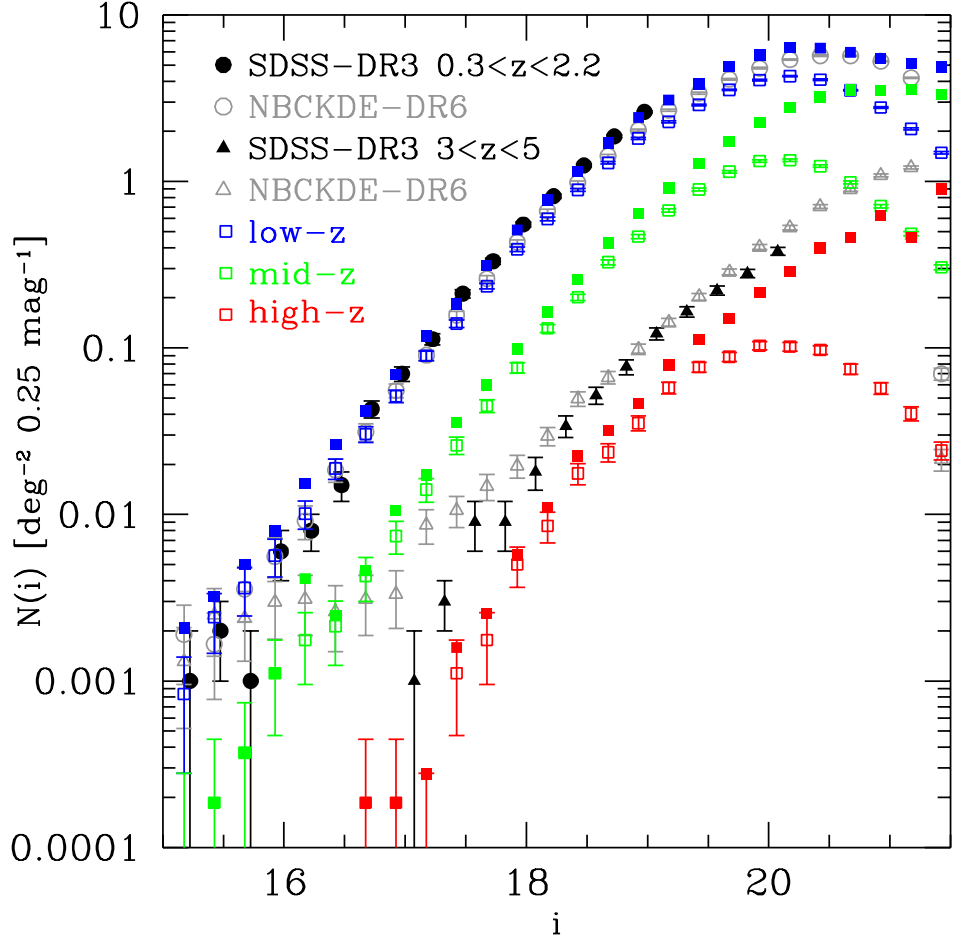}
\caption{Quasar number counts as a function of redshift and $i$-band magnitude.  Black and gray points, respectively give the spectroscopic and photometric number counts as reported in \citet[e.g., Fig.~9][]{rmg+09}; circles for $z<2.2$ and triangles for $3<z<5$.  The open blue, green, and red squares give the raw number counts (with 1-$\sigma$ Poisson error bars) for the candidates reported herein.  The filled colored squares give the number counts corrected using Figure~\ref{fig:fig15}.  The mid-$z$ and high-$z$ samples bracket the redshift space of the old $3<z<5$ sample, but show no sign of the contamination at the bright end (flattening of the number counts) seen in the old sample.
\label{fig:fig14}}
\end{figure}

In Figure~\ref{fig:fig14} we reproduce Figure~9 from
\citet{rmg+09} which showed both the spectroscopic and photometric
quasar number counts in two redshift ranges.  Here we have overplotted
the number counts of our quasar candidates selected as low-$z$,
mid-$z$, and high-$z$ candidates.  

\begin{figure}[h!]
\epsscale{0.6}
\plotone{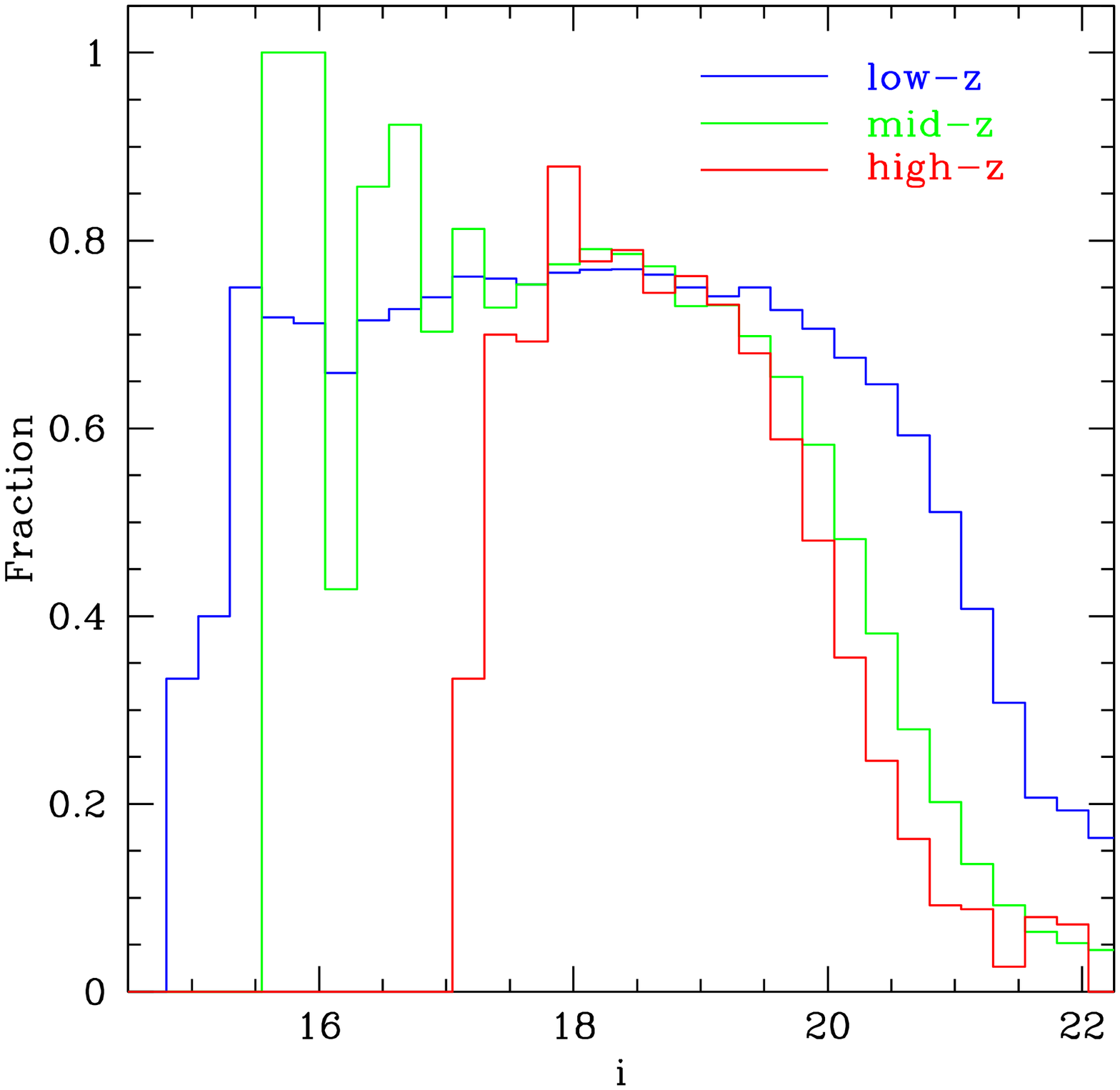}
\caption{Ratio of objects in the master catalog objects recovered by our algorithm  to the master quasar catalog (prior to matching to mid-IR photometry).  This corrects for objects too faint in the mid-IR to match the optical, objects rejected from the IR catalog due to flags, the exclusion of extended sources, and the incompleteness of the selection algorithm itself.  The fraction is given as a function of $i$-band magnitude in the three redshift ranges we have explored (low-$z$: blue, mid-$z$: green, high-$z$: red).
  \label{fig:fig15}}
\end{figure}

In this figure, open points represent the raw number counts, while the
closed points give the completeness-corrected number counts.  As we
will do for the luminosity function analysis below, the objects going
into the raw number counts presented here are limited to those in the
SDSS ``legacy area'' (area = 10778.306\,deg$^2$) and are classified as
either quasars or unknown.  The unknown sources are restricted to
robust candidates as defined above in Section~\ref{sec:robust}
The completeness corrections for this sample are given by the fraction
of master quasar catalog objects recovered by our algorithm with these
constraints as shown in Figure~\ref{fig:fig15}.  This analysis
converts the raw counts to the total number of quasars expected
(accounting for incompleteness of the selection algorithm, lack of
mid-IR photometry, non-stellar morphology, and flag-rejection).

We specifically find that the corrected low-$z$ number counts are a
good match to the spectroscopic number counts at $i\sim17$, being
somewhat incomplete at $i\sim19$ (but probing to $i\sim20.5$), and
exhibiting perhaps a factor of two contamination in the brightest bin
shown.  For mid-$z$ quasars our sample appears to be filling in the
gap in the SDSS selection over $19.1<i<20.2$, while exhibiting less
contamination than our previous photometric sample (as evidenced by a
lack of a plateau at the bright end).  The high-$z$ number counts do
not show any obvious sign of contamination from bright stars (once we
have imposed the restrictions noted above).

These number counts are thus consistent with our new catalog being
both relatively complete (to within a deterministic correction) and
efficient.  If the efficiency was low (and thus the contamination was
high), we would expect significant deviations from the slopes of the
spectroscopic number counts.  We see none of the excess in our current
photometric sample as we saw in the high-$z$ sample from
\citet{rmg+09} and the faint-end counts are consistent with the
optical+infrared selected candidates from \citet[][Fig.~12]{rdl+09},
which performed a selection similar to our current selection, but over
a much smaller area of sky ($\sim$24\,deg$^2$).

\begin{figure}[h!]
\epsscale{0.6}
\plotone{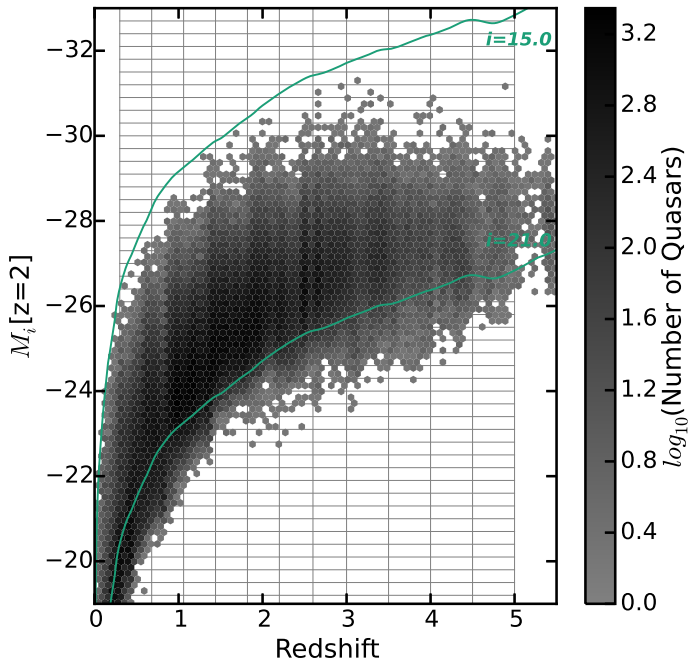}
\caption{Absolute magnitude (luminosity) vs. redshift for our photometric quasar candidates with number of objects displayed as gray-scale hex bins.  Teal lines indicate the bright limit of SDSS and the adopted limiting magnitude of our QLF analysis.    The light gray grid lines delineate the bins used to compute the QLF in Figure~\ref{fig:fig17}.  
\label{fig:fig16}}
\end{figure}

\begin{figure}[h!]
\plotone{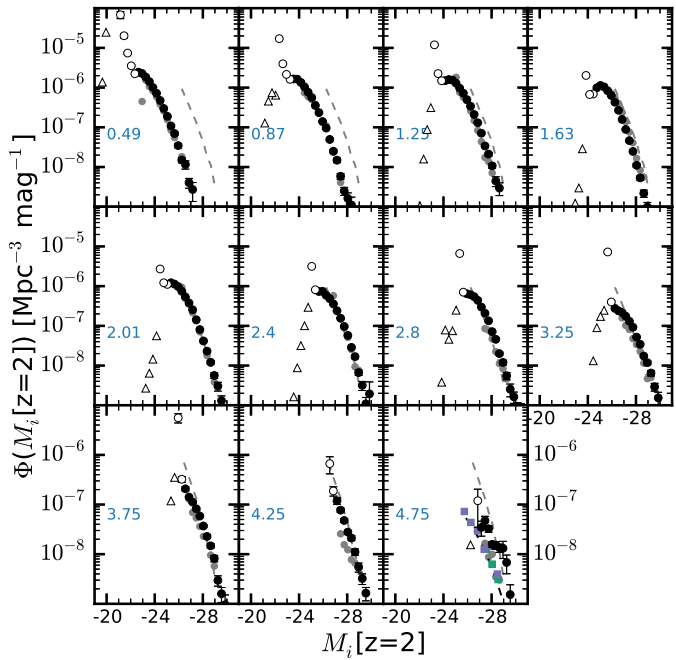}
\caption{Quasar luminosity function in 11 redshift bins.  Filled black circles are photometric objects from our catalogs brighter than the adopted limiting magnitude.  Open circles are those where the limiting magnitude cuts though the ($L,z$) bin and thus have uncertain corrections (error bars are Poisson only), while the open triangles indicate (uncorrected) lower limits.  Grey points are at the spectroscopic QLF values from \citep{rsf+06}, where the dashed grey line repeats the spectroscopic QLF from $z=2$ in each redshift panel.  In the $z=4.75$ panel, we overplot the data (purple and teal) and best fit (dashed black line) from \citet{mjf+13}.   The photometric QLF matches the spectroscopic QLF quite well, especially considering that this was not one of the goals of this investigation.  The excess density at $z=4.25$ indicates either an under-correction of the completeness by \citet{rsf+06} or contamination in our sample---likely a combination of both.
\label{fig:fig17}}
\end{figure}

While our goal in this work was not to determine the luminosity
function of quasars, but rather to take the next step in creating
optimal photometric catalogs of quasars, it is nevertheless useful to
examine the quasar luminosity function (QLF) as determined from our
catalog.  In Figure~\ref{fig:fig16} we show the absolute magnitude
(luminosity) and photometric redshift distribution of our data using
the same redshift and luminosity bins as \citet{rsf+06} and we compare
the resulting luminosity function in these bins in
Figure~\ref{fig:fig17}.  We have taken the limiting magnitude to be
$i<21$ as shown, since that is where our completeness falls below 50\%
according to Figure~\ref{fig:fig3}.  However, there is no single
limiting magnitude for this investigation as we have simply matched
all of the SDSS optical sources with MIR sources from {\em WISE} and
{\em Spitzer}.  Objects fainter than $i=21$ can be included in the
catalog if they are bright enough in the MIR, but they are excluded
from our main QLF analysis.  The gradient in the density of points
near the $i=21$ limit in Figure~\ref{fig:fig16} might suggest that we
are complete to deeper than this limit at low-$z$, but also that the
completeness is at a somewhat brighter magnitude high-$z$.

To produce the QLF results shown in Figure~\ref{fig:fig17} we
restricted the catalog using the same cuts as above for the number
counts, namely limiting to known quasars and ``robust'' unknown
sources, both within the legacy area.
In this presentation we make two corrections to the raw data.  First,
we correct for incompleteness as a function of $i$-band magnitude and
redshift by weighting by the fraction of training set quasars
recovered by our algorithm.  Next we correct the photometric redshifts
by weighting each object by the ratio of the number of spectroscopic
redshifts to photometric redshifts for our training set quasars.  That
is, if there were really 100 spectroscopic training-set quasars at
$z=1.45$--$1.55$, but the photometric redshift estimates for those
quasars placed 120 quasars in the same bin, then we would weight each
new photometric quasar candidate in that photo-$z$ bin by $100/120$.

We find reasonable agreement with the spectroscopic QLF points of
\citet{rsf+06} given that the focus of this work was not the rigorous
computation of the QLF.  Specifically the black points in
Figure~\ref{fig:fig17} are in good agreement with the SDSS points
(gray) down to the flux limit of SDSS and appear to be well-behaved
another magnitude deeper than the SDSS data.

An exception is the deviation from \citet{rsf+06} seen in the $z=4.25$
panel, where our photometric sample has a space density that is a
factor of a few higher than SDSS at $M_i(z=2)\sim-27$.  This is likely
to be caused either by contamination from non-quasars in our sample or
under-correction of the SDSS completeness in this redshift range.  If
it is incompleteness, the origin may be a greater sensitivity of our
method to dust-reddened (but unobscured) quasars.  Indeed,
\citet{lrs+15}, using an MIR-selected sample, similarly find a high
fraction of redder quasars at high redshift.  Interestingly, this
high-redshift QLF exhibits a steeper QLF slope than \citet{rsf+06},
and is more consistent with the results of \citet{jfa+08},
\citet{rmw+13}, and \citet{mjf+13}.  At $z=4.75$ the errors are
somewhat larger, but the QLF is broadly consistent with
\citet{mjf+13}.

\subsection{Future}

One of the goals of this work is to set the stage for next-generation
clustering investigations using high-redshift quasars.  The SDSS
quasar sample lacks sufficient density to test the luminosity
dependence of quasar clustering \citep{lhc+06} such as proposed by
\citet{hlh+07}.  For example the work by \citet{sso+07} used a sample
of only $\sim4000$ quasars at $2.9<z<5.4$ over $\sim4000$\,deg$^2$.
Here we cover more than double that area and nearly double the sample
size, but over an even smaller redshift range.  The various optical
and MIR deep fields would enable the discovery of more objects by
probing much deeper, but they are limited in their utility for
high-$z$ clustering investigations by their small area and the MIR
bias against high-$z$ quasars.

Substantial gains should come from pairing this method with the data
coming from the SpIES project (Timlin, Ross, Richards et al.\ 2015),
which has just completed tiling $\sim125$\,deg$^2$ of the SDSS Stripe
82 region \citep[e.g.,][]{Annis11}.  We can estimate the number of
high-$z$ quasars in the SpIES area from the SWIRE ELAIS-N2 field
(4.2\,deg$^2$) which has the same depth as SpIES (but has not been
covered by SERVS).  In that field we find 32 high-$z$ quasar
candidates, 24 of which appear to have robust photometric redshifts.
Thus we predict that SpIES will contain of order 5--7 high-$z$ quasars
per square degree or a total of 625--875 objects.  This density should
be sufficient for powerful tests of the clustering of quasars as a
function of luminosity at high redshift.

This work is further a proof of concept for future quasar surveys
using both ground- and space-based data, such as could be done by
combining photometric data from Pan-STARRS \citep{kab+02} ($grizy$),
SkyMapper \citep{SkyMapper} ($uvgriz$), the Dark Energy Survey
\citep{des05} ($grizY$), Hyper
Suprime-Cam\footnote{http://www.naoj.org/Projects/HSC/surveyplan.html}
($grizy$), the Large Synoptic Survey Telescope \citep{lsst08}
($ugrizy$), the NEOWISE extension to the {\em WISE} program
\citep{NEOWISE} (using the two shortest {\em WISE} bandpasses), {\em Euclid}
\citep{Euclid} ($YJH$) or for future spectroscopic programs like the
Dark Energy Spectroscopic Instrument (DESI; \citealt{saa+11}).  We
have shown that using the combination of optical and MIR photometry is
better (for unobscured quasars) than either data set alone and that
there are considerable gains to be made from the use of modern
statistical methods in performing multi-dimensional selection.

\section{Conclusions}
\label{sec:conclusions}

Using a proven kernel density estimation technique, we identify
885,503 type 1 quasar candidates within the imaging footprint of the
Sloan Digital Sky Survey by combining the SDSS optical data with
mid-IR imaging from {\em WISE} and {\em Spitzer}.  Among these objects
are 6779 robust, $3.5<z<5$ quasar candidates that have no previous
spectroscopic or photometric classification.  This increase is
possible due to incompleteness of MIR-only color selection in this
redshift range and the difficulty of variability selection for faint,
high-redshift quasars, and offers an opportunity to expand our
exploration of the high-redshift universe.

The optical and MIR color distributions shown in
Figure~\ref{fig:fig8} and \ref{fig:fig9} are good
matches to the distributions of the training set quasars, but extend
to fainter limits in both the optical and MIR.  They also clearly
demonstrate an increased completeness to high-redshift quasars
(particularly at $3.5<z<5$ where MIR color selection is incomplete due
to spectral features pushing the colors of these objects bluer than
typical MIR color-cuts).  

Photometric redshift estimates of these candidates using optical and
MIR photometry are accurate to $\Delta z \pm 0.3$ at least 83\% of the
time, improving to 93\% where there also exists near-IR photometry;
see Figure~\ref{fig:fig10}.  Comparison with the known colors of
objects at the expected redshift (Figure~\ref{fig:fig11}) can help to
identify potential contaminants and/or those objects with erroneous
photo-$z$.

Our new candidates even include robust targets within the well-covered
COSMOS and Bo\"{o}tes fields, where an increased density of
spectroscopic quasars would aid in clustering and absorption line
studies.  This includes over 50 robust, new high-$z$ quasar candidates in
both of the fields (where there exists deeper-than-average MIR photometry).

Generally our algorithm is simply finding quasars that are fainter
than the SDSS spectroscopic limits, and
that should not necessarily have received
SDSS spectroscopic followup.  However, there are a number of bright
low-$z$ candidates without SDSS spectroscopy that are likely to be
low-luminosity AGNs rather than luminous quasars.
Figures~\ref{fig:fig12} and \ref{fig:fig13} present the
magnitude and expected redshift distributions of both the new
candidates and the known quasars.

We are able to explore the completeness and contamination of the
method using number counts and luminosity function analysis.
Figure~\ref{fig:fig14} demonstrates that our algorithm is
relatively complete to known low-$z$ quasars (accounting for our
restriction to optical point sources) and shows no obvious sign of
contamination from bright stars at any redshift.  The QLF shown in
Figure~\ref{fig:fig17} agrees well with the results from SDSS
\citep{rsf+06}, but suggest a steeper slope to the QLF at high-$z$
(consistent with \citealt{mjf+13}) and may be more sensitive to
dust-reddened quasars.  Future work will expand that presented herein
by incorporating more information (variability, proper motion, etc.)
and using survey data that probes deeper in the optical.

\acknowledgments

GTR and ADM acknowledge the generous support of a research fellowships
from the Alexander von Humboldt Foundation at the Max-Planck-Institut
f\"{u}r Astronomie and GTR is grateful for the hospitality of the
Astronomisches Rechen-Institut.  This work was supported in part by
NASA-ADAP grant NNX12AI49G and NSF grant 1411773.  ADM acknowledges
support from NASA ADAP grant NNX12AE38G and NSF grant 1211112.

Funding for the SDSS and SDSS-II has been provided by the Alfred
P. Sloan Foundation, the Participating Institutions, the National
Science Foundation, the U.S. Department of Energy, the National
Aeronautics and Space Administration, the Japanese Monbukagakusho, the
Max Planck Society, and the Higher Education Funding Council for
England. The SDSS is managed by the Astrophysical Research Consortium
for the Participating Institutions. The Participating Institutions are
the American Museum of Natural History, Astrophysical Institute
Potsdam, University of Basel, Cambridge University, Case Western
Reserve University, University of Chicago, Drexel University,
Fermilab, the Institute for Advanced Study, the Japan Participation
Group, Johns Hopkins University, the Joint Institute for Nuclear
Astrophysics, the Kavli Institute for Particle Astrophysics and
Cosmology, the Korean Scientist Group, the Chinese Academy of Sciences
(LAMOST), Los Alamos National Laboratory, the Max-Planck-Institute for
Astronomy (MPIA), the Max-Planck-Institute for Astrophysics (MPA), New
Mexico State University, Ohio State University, University of
Pittsburgh, University of Portsmouth, Princeton University, the United
States Naval Observatory, and the University of Washington.

Funding for SDSS-III has been provided by the Alfred P. Sloan
Foundation, the Participating Institutions, the National Science
Foundation, and the U.S. Department of Energy Office of Science. The
SDSS-III web site is http://www.sdss3.org/.  SDSS-III is managed by
the Astrophysical Research Consortium for the Participating
Institutions of the SDSS-III Collaboration including the University of
Arizona, the Brazilian Participation Group, Brookhaven National
Laboratory, Carnegie Mellon University, University of Florida, the
French Participation Group, the German Participation Group, Harvard
University, the Instituto de Astrofisica de Canarias, the Michigan
State/Notre Dame/JINA Participation Group, Johns Hopkins University,
Lawrence Berkeley National Laboratory, Max Planck Institute for
Astrophysics, Max Planck Institute for Extraterrestrial Physics, New
Mexico State University, New York University, Ohio State University,
Pennsylvania State University, University of Portsmouth, Princeton
University, the Spanish Participation Group, University of Tokyo,
University of Utah, Vanderbilt University, University of Virginia,
University of Washington, and Yale University.

This work is based [in part] on archival data obtained with the
Spitzer Space Telescope, which is operated by the Jet Propulsion
Laboratory, California Institute of Technology under a contract with
NASA. Support for this work was provided by an award issued by
JPL/Caltech (\#1378034).  This publication makes use of data products
from the Wide-field Infrared Survey Explorer, which is a joint project
of the University of California, Los Angeles, and the Jet Propulsion
Laboratory/California Institute of Technology, funded by the National
Aeronautics and Space Administration.  This work has made extensive
use of the TOPCAT \citep{tay05} and STILTS \citep{tay06} tools.  We
thank Joe Hennawi, Nadia Zakamska, Alex Gray, and the anonymous
referee for their contributions.



{\it Facilities:} \facility{Sloan, Spitzer, WISE}.

\bibliography{ms,../../sdsstech,../../richards_shortnames}
\bibliographystyle{apj3}


\end{document}